\documentclass[preprint, aps, amsmath,showkeys, nofootinbib, superscriptaddress]{revtex4-2}
\usepackage{graphicx}
\usepackage{amsmath}
\usepackage{booktabs}
\usepackage{multirow}
\usepackage{array}
\usepackage{longtable}
\usepackage{ulem}
\usepackage{xcolor}

\begin{document}


\title{Antikaon absorption in the nuclear medium: the role of hadron self-energies and implications for kaonic atoms}


\author{J. \'{O}bertov\'{a}}
\email[]{jaroslava.obertova@fjfi.cvut.cz}
\affiliation{Faculty of Nuclear Sciences and Physical Engineering, Czech Technical University in Prague, B\v{r}ehov\'{a} 7, 115 19, Prague, Czech Republic}

\author{\`{A}. Ramos}
\affiliation{Departament de Física Quàntica i Astrofísica, Universitat de Barcelona,  Martí i Franquès, 1, 08028 Barcelona, Spain}
\affiliation{Institut de Ciències del Cosmos, Universitat de Barcelona, Martí i Franquès, 1, 08028 Barcelona, Spain}

\author{J. Mare\v{s}}
\affiliation{Nuclear Physics Institute of the Czech Academy of Sciences, 25068 \v{R}e\v{z}, Czech Republic }


\date{\today}

\begin{abstract}

A systematic study of all relevant in-medium effects on the total $K^-$-nuclear potential is presented in this work. The $K^-N$ scattering amplitudes, including Pauli blocking effects and hadron self-energies (hyperons, nucleons, pions and kaons), are derived within a next-to-leading order chiral meson-baryon coupled-channel interaction model. These amplitudes are employed in a microscopic model of the $K^-$-nuclear potential in symmetric nuclear matter that includes one-, two- and, when the kaons and pions are dressed, also multinucleon absorption processes.  The potential is then applied in calculations of the strong energy shifts and widths of 64 measured kaonic atom levels. The comparison of the results of the full model that includes Pauli correlations and hadron self-energies with data provides $\chi^2 /d.p=1.5$, the lowest value obtained by a theoretical model to date and comparable with that of the best fitted phenomenological potentials. Furthermore, the calculated branching ratios for mesonic and non-mesonic absorption channels in kaonic carbon and kaonic neon are in good agreement with available data.

\end{abstract}


\maketitle

\newpage
\section{Introduction}

Kaonic atoms provide essential information about the $K^-$-nuclear potential below threshold. The data set extracted from bubble chamber experiments \cite{friedman:1994npa, friedman:2007pr} span a region of nuclei from $^{7}$Li up to $^{238}$U and cover 64 data points for strong energy shifts and widths. The recent experiment E62 at the J-PARC facility delivered new high precision data on energy shifts and widths in kaonic $^{3}$He and $^{4}$He atoms \cite{jparc}. The SIDDHARTA-2 collaboration measured the energy shift and width in kaonic deuterium with the highest possible precision and the data are currently being analyzed \cite{Sgaramella:SIDDHARTA2}. Additional information about the subthreshold interaction of $K^-$ with nucleons comes from the analyses of $\pi\Sigma$ spectra in the region of the $\Lambda(1405)$ resonance \cite{HADES, CLAS1, CLAS2}.

 Global analysis of kaonic atom data using the density-dependent optical potential \cite{batty:1997pr, friedman:2007pr} yielded a deeply attractive $K^-$-nuclear potential at saturation density $\rho_0$, Re$V_{K^-}(\rho_0) \approx -(150 - 200)$~MeV, and also fairly absorptive. Recent phenomenological analysis of kaonic atom data by J. Yamagata-Sekihara et al. \cite{Yamagata-Sekihara:PTEP2025}, including the latest measurement of kaonic $^{3}$He and $^{4}$He from J-PARC, confirmed the conclusions about the depth of the phenomenological $K^-$-nuclear potential from previous analyses \cite{batty:1997pr, friedman:2007pr}. They found that two sets of solutions for the $K^-$ potential at saturation density, one with deeply attractive and shallow absorptive parts and the other with shallow attractive and deeply absorptive parts, describe the new data on kaonic $^{3}$He and $^{4}$He equally well. However, it was pointed out \cite{fgNPA2017} that kaonic atom data are not sensitive to the region of saturation density and can reliably constrain the total $K^-$-nuclear potential only up to $\approx 0.5\rho_0$. 

Theoretical models of the low-energy $\bar{K}N$ interaction based on the chiral SU(3) meson-baryon effective Lagrangian were developed by various groups over the last two decades \cite{pnlo, kmnlo, b, m, feijoo:2019prc}. The parameters of the chiral models are tuned to reproduce the low-energy $K^-p$ scattering data \cite{kp_crosssection1, kp_crosssection2, kp_crosssection3}, threshold branching ratios \cite{kp_ratios1, kp_ratios2}, and the precise strong energy shift and width in kaonic hydrogen \cite{SIDDHARTA}. The data fix mainly the $K^-p$ amplitude at and above threshold. The $K^-n$ scattering amplitude is strongly model dependent even at and above threshold due to the lack of experimental data. However, the measurement of kaonic deuterium by the SIDDHARTA-2 Collaboration \cite{SIDDHARTA2:2024nc, SIDDHARTA2:2024jinst} is expected to provide important constraint also on the $K^-N$ interaction in the isospin 1 channel.

The first estimates of the $K^-$ potential in a nuclear medium based on the  self-consistent calculation within the chiral models gave relatively shallow Re$V_{K^-}(\rho_0)$, in the range $ \approx -(40 - 60)$~MeV \cite{ramos:2000npa, Cieply:2001npa, Koch:2000npa}. Ramos and Oset \cite{ramos:2000npa} studied the properties of $K^-$ in the nuclear medium using the leading-order chiral $K^-N$ interaction \cite{Oset:1997it}, a model that also reproduced well the pole position of $\Lambda(1405)$. They derived the $K^-$-nuclear optical potential that included the Pauli blocking effect and the dressing of $K^-$ and pions in the medium. Similar self-consistent calculations of antikaons and hyperons in matter within a relativistic framework were developed in Ref.~\cite{Lutz:2001dq}, improving over earlier works with only Pauli blocking effects \cite{Lutz:1998plb,Lutz:1994npa}.
 The microscopic $K^-$-nuclear potential obtained in Ref.~\cite{ramos:2000npa} was then applied in the calculations of kaonic atoms \cite{Hirenzaki:2000da, bacaNPA2000} and achieved a $\chi^2$/d.p = 3.8 \cite{bacaNPA2000}. Later on, the chiral model of Ref.~\cite{Oset:1997it} was extended to next-to-leading order \cite{feijoo:2019prc} and denoted as the Barcelona (BCN) model.

A study of $K^-$-nuclear potentials based on the state-of-the-art chiral models confirmed the depth of Re$V_{K^-}(\rho_0) \approx -(30 - 110)$~MeV and very shallow imaginary part Im$V_{K^-}(\rho_0) \approx - (10 - 20)$~MeV  \cite{hmPRC2017}. However, Friedman and Gal \cite{fgNPA2017} demonstrated that the $K^-$-nuclear potentials constructed from chiral $\bar{K}N$ amplitudes are generally unable to fit the kaonic atom data. 
The reason is that the chiral models consider only mesonic decay channels of $K^-$ in the medium, $K^-N \rightarrow \pi Y$ with $Y=\Sigma, \Lambda$. An important part of the $K^-$ interaction in the nuclear medium are absorption processes on 2 and more nucleons, mainly the non-mesonic decay channel $K^-NN \rightarrow YN$. The measurement of the multinucleon absorption fraction in bubble chamber experiments \cite{vander:1977nc, moulder:1971npb, davis:1968nc, katz:1970prd} revealed that multinucleon processes amount to approximately 20\% of all $K^-$ absorptions in the surface region of atomic nuclei. Recent measurement of $K^-$ two-nucleon branching ratio in analysis of $\Lambda p$ and $\Sigma p$ spectra of antikaon absorption in $^{12}$C by the AMADEUS collaboration \cite{amadeus19, amadeus2020} confirmed the value reported by old bubble chamber experiments.

The analysis in Ref.~\cite{fgNPA2017} showed that only the Prague \cite{pnlo}, Kyoto-Munich \cite{kmnlo}, and BCN \cite{feijoo:2019prc} chiral models are able to reproduce simultaneously kaonic atom data and single-nucleon absorption fractions when supplemented by a phenomenological density-dependent multinucleon potential. Compatibility of these three models with the data was confirmed by the AMADEUS collaboration, which measured the modulus of the non-resonant $K^-n \rightarrow \Lambda \pi^-$ production amplitude about 33 MeV below the $\bar{K}N$ threshold, in absorption of $K^-$ on the $^{4}$He target \cite{amadeus2018}. The BCN model was recently employed in a calculation of the $K^-p$ correlation function \cite{Encarnacion:2025prd} and agrees with the femtoscopic ALICE data \cite{ALICE:kp_femto} within $2\sigma$.

The first calculation of the $K^-NN$ absorption fraction in a nuclear medium within the chiral model was done by Sekihara et al. \cite{Sekihara:2012prc}. In Ref. \cite{hrtankova:2020prc}, we developed a microscopic model for the $K^-N$ and $K^-NN$ absorption in nuclear medium adopting the chiral $K^-N$ amplitudes derived within the BCN model. Unlike in Ref.~\cite{Sekihara:2012prc}, we considered the in-medium modification of the free-space $K^-N$ amplitudes due to the Pauli principle and derived the total $K^-$ optical potential as a function of nuclear density. The model successfully reproduced the ratio of $\Lambda p$ and $\Sigma^0 p$ final states in $K^-$ two-nucleon absorption processes on $^{12}$C \cite{hrtankova:2020prc, amadeus19}. The microscopic $K^-N+K^-NN$ potential was applied in calculations of strong energy shifts and widths in kaonic atoms throughout the periodic table \cite{ofmPRC2022}. Direct comparison of the calculation with 64 data points showed a significant improvement in $\chi^2$ when the two-nucleon absorption processes were taken into account. However, in order to get $\chi^2$ as low as for the $K^-N +$phenomenological multinucleon potential, it was still needed to add an additional phenomenological term accounting for $K^- - 3N(4N)$ processes.

In the present study, we properly include hadron ($Y, N, K^-, \pi$) self-energies in the BCN $K^-N$ chiral in-medium amplitudes as well as in the $K^-NN$ absorption model. We perform a systematic study of the influence of the various hadron self-energies on the energy dependence of the $K^-N$ chiral amplitudes and, consequently, on the depth of the total $K^-$ potentials in symmetric nuclear matter. The potentials are then applied in calculations of energy shifts and widths in kaonic atoms, as well as in evaluation of mesonic and non-mesonic branching ratios. The full potential model, including Pauli correlations in the medium, all hadron self-energies, and multinucleon absorption processes, describes for the first time the data as good as the best fit provided by the $K^-N$+phenomenological multinucleon potential based on the BCN chiral amplitudes.

The paper is organized as follows. In Sec.~\ref{KNamplitudes}, we briefly describe the BCN chiral amplitudes and their modification due to the hadron self-energies. Then, in Sec.~\ref{KNNmodel}, we present the $K^-N$ and $K^-NN$ absorption model including hadron-self energies and construct the total $K^-$ potential in symmetric nuclear matter. In Sec.~\ref{atoms}, we discuss the application of the $K^-$ potential and corresponding subthreshold kinematics in the calculation of kaonic atoms. In Section \ref{results}, we present the evolution of the total $K^-$ optical potential in nuclear matter, obtained with step-by-step inclusion of the hadron self-energies. Next, we present calculated energy shifts and widths in kaonic atoms and mesonic and non-mesonic branching ratios in kaonic carbon and kaonic neon compared with available experimental data. In the last Section \ref{conclusions}, we present a summary of our results.

\section{Model}
\label{model}

\subsection{$K^-N$ amplitudes with hadron self-energies}
\label{KNamplitudes}

In this section we describe how the $K^-N$ scattering amplitudes are modified in the nuclear medium. Throughout this work we employ the $K^- N$ model developed in Ref.~\cite{feijoo:2019prc}, which builds the interaction from a chiral SU(3) Lagrangian up to next-to-leading order (NLO) and implements unitarization in coupled channels. More specifically, this model, referred here as the BCN model, obtains the $K^-p$ scattering amplitude and related channels from the solution of the on-shell Bether-Salpeter equation:
\begin{equation}
    t_{ij}=v_{ij} + v_{ik} G_k t_{kl} \ ,
    \label{eq:BS}
\end{equation}
with $i,j,k=\{K^- p$, $\bar{K}^0 n$, $\pi^0 \Lambda$, $\pi^0\Sigma^0$, $\pi^-\Sigma^+$, $\pi^+\Sigma^-$, $\eta \Lambda$, $\eta\Sigma^0$, $K^+\Xi^-$, $K^0\Xi^0\}$, where the interaction
\begin{equation}
    v_{ij} = v^{\rm WT}_{ij} + v^{\rm D}_{ij}  + v^{\rm C}_{ij}  + v^{\rm NLO}_{ij} 
\end{equation}
contains the contribution of the contact (Weinberg-Tomozawa) term, the direct (D) and crossed (C) Born terms, and the NLO term.  The loop function $G_k$ stands for the meson-baryon propagator
\begin{equation} \label{Loop_integral}
G_k={\rm i}\int \frac{d^4q_k}{{(2\pi)}^4}\frac{2m_{B_k}}{{(P-q_k)}^2-m_{B_k}^2+{\rm i}\epsilon}\frac{1}{q_k^2-m_{M_k}^2+{\rm i}\epsilon} ,
\end{equation} 
with $m_{B_k}$ and $m_{M_k}$ denoting the mass of the baryon and meson in the $k^{\rm th}$ channel, respectively.
Its logarithmic divergence is handled by applying a dimensional regularization scheme, giving rise to
\begin{eqnarray}
    G_{k}^{\rm DR}&=&\frac{2m_{B_k}}{16\pi^2}\Big\{ a_k(\mu)+\ln\frac{m_{B_k}^2}{\mu^2}+\frac{m_{M_k}^2-m_{B_k}^2+s}{2s}\ln\frac{m_{M_k}^2}{m_{B_k}^2}+ \nonumber \\ 
  &+&\frac{q_k}{\sqrt{s}}\left[\ln\left(s-(m_{B_k}^2-m_{M_k}^2)+2q_k\sqrt{s}\right)~+~\ln\left(s+(m_{B_k}^2-m_{M_k}^2)+2q_k\sqrt{s}\right) \right. \nonumber \\
  &&\phantom{~~~~}\left.-\ln\left(-s+(m_{B_k}^2-m_{M_k}^2)+2q_k\sqrt{s}\right)-\ln\left(-s-(m_{B_k}^2-m_{M_k}^2)+2q_k\sqrt{s}\right) \right] \Big\} \ , 
  \label{eq:GmatrixDR}
\end{eqnarray}
where the subtraction constants $a_k$, which replace the divergence for a given regularization scale $\mu$ (chosen to be 1 GeV), were fitted, together with the low-energy constants of the chiral Lagrangian, to a large set of experimental data (see Ref.~\cite{feijoo:2019prc} for details).
One can also employ a cut-off scheme to regularize the loop function,
\begin{eqnarray}
G_{k}^\Lambda(\sqrt{s}) &=&  \int_{\mid {\vec q} \mid < \Lambda} \, \frac{d^3 q}{(2
\pi)^3} \,
\frac{1}{2 \omega_k (\vec q\,)}
\,
\frac{m_{B_k}}{E_k (-\vec{q}\,)} \,
\frac{1}{\sqrt{s}- \omega_k (\vec{q}\,) - E_k (-\vec{q}\,) + i
\epsilon} \ ,
\label{eq:gprop}
\end{eqnarray}
with $E_k $ and $\omega_k$ being the energy of the baryon and meson in the $k^{\rm th}$ channel, respectively.
This expression, adopted in the early Oset-Ramos (OR) model \cite{Oset:1997it}, is more practical for implementing the in-medium effects on the scattering amplitudes. 

\begin{figure}[h!]
    \centering
    \includegraphics[width=0.45\textwidth]{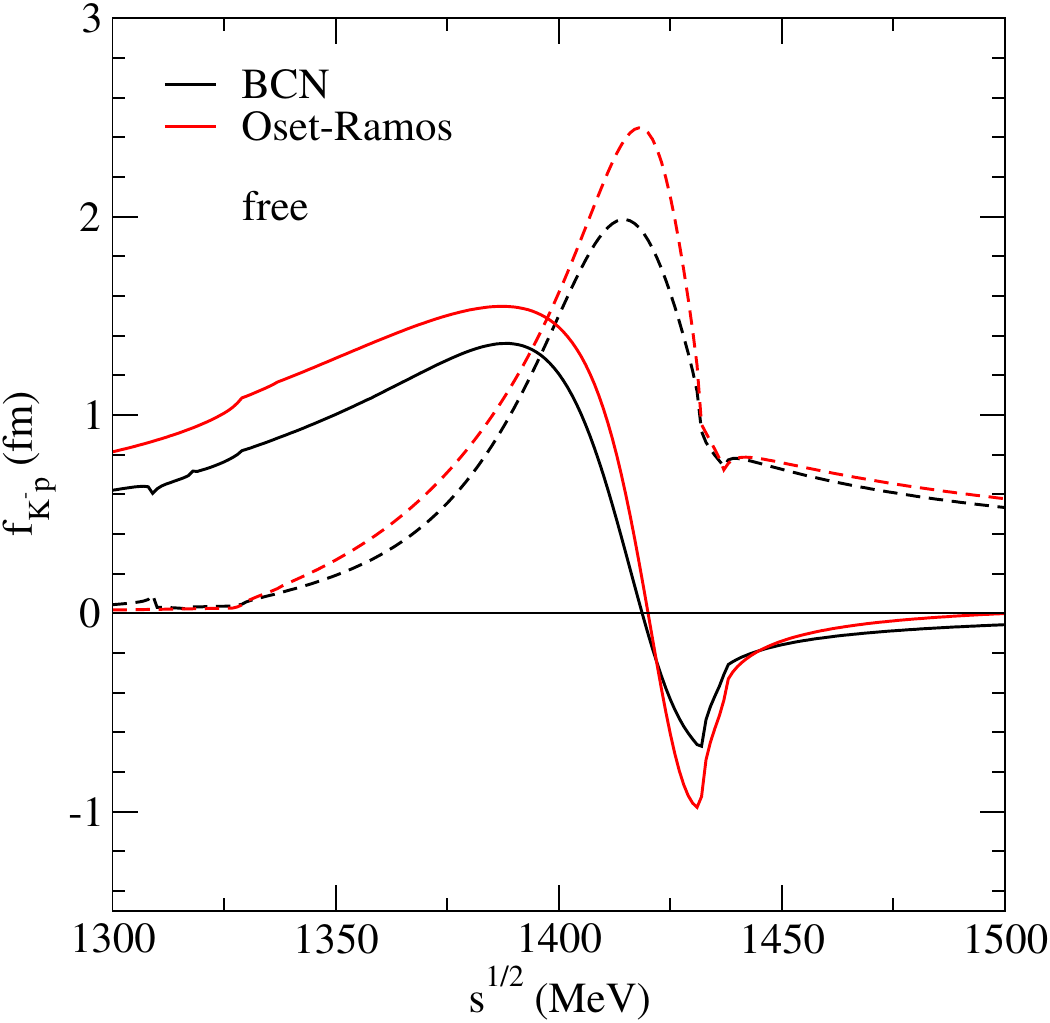}~~~~~~
\includegraphics[width=0.455\textwidth]{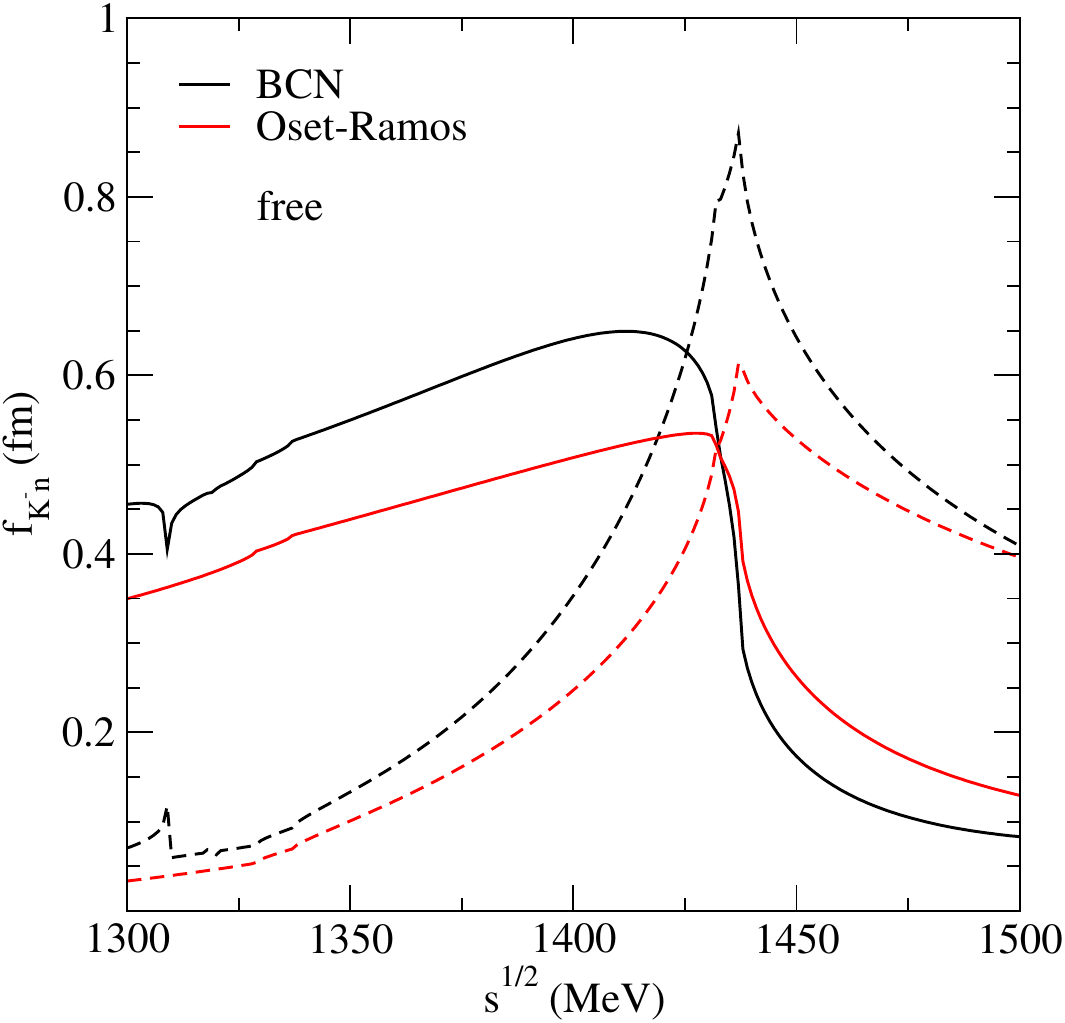} 
    \caption{Real part (solid lines) and imaginary part (dashed lines) of the $K^-p\rightarrow K^-p$ free-space amplitude, $f_{K^-p}$ (left panel),  and the $K^-n\rightarrow K^-n$ free-space amplitude, $f_{K^-n}$ (right panel), for the BCN model (black lines) and the Oset-Ramos model (red lines).}
    \label{fig:0}
\end{figure}

In Fig.~\ref{fig:0}, we present a comparison of the energy dependence of the scattering amplitudes $f_{K^- p}$ (left panel) and $f_{K^-n}$ (right panel) in free space,  evaluated in the leading-order OR model \cite{Oset:1997it} and in the next-to-leading order BCN \cite{feijoo:2019prc} model, defined from the t-matrix elements as:
\begin{equation}
f_{K^-N} = -\frac{m_N}{4\pi\sqrt{s}}t_{K^- N \rightarrow K^- N}    
\end{equation}
with $N=p,n$. The $K^- n$ amplitude has been obtained from the amplitudes in the charge zero sector as $t_{K^- n \rightarrow K^- n} = (t_{K^- p \rightarrow K^- p} + 
t_{{\bar K}^0 n \rightarrow {\bar K}^0 n} -2t_{K^- p \rightarrow {\bar K}^0 n})/2$, invoking isospin symmetry considerations. Both OR and BCN models yield the same energy dependence of the $K^-N$ scattering amplitudes. The $K^-p$ amplitude below threshold becomes less attractive and less absorptive in the BCN model compared to the OR model. On the other hand, the trend for the $K^-n$ amplitude is opposite.

To incorporate the in-medium effects on the amplitudes, we follow the procedure of Ref.~\cite{ramos:2000npa}, which we describe here briefly for completeness. The Pauli principle,
preventing the intermediate nucleon states to lie below the
Fermi momentum, $p_F$, is a clear source of density dependence.
This is implemented upon replacing the free nucleon
propagator by the in-medium one in the loop integral of Eq.
(\ref{eq:gprop}), which then becomes:
\begin{eqnarray}
G_{k}^\Lambda(P^0,\vec{P},\rho)
&=& \int_{\mid {\vec q} \mid < \Lambda} \, \frac{d^3 q}{(2
\pi)^3} \,
\frac{1}{2 \omega_k (\vec q\,)}
\,
\frac{m_{B_k}}{E_k (-\vec{q}\,)} \,
\left\{
\frac{1 - n(\vec{q}_{\rm lab})}{\sqrt{s}- \omega_k (\vec{q}\,) -
E_k (-\vec{q}\,) + i
\epsilon} \right. \nonumber \\
& &
\phantom{\int_{\mid {\vec q} \mid < q_{\rm max}} \, \frac{d^3
q}{(2
\pi)^3} \,
\frac{1}{2 \omega_k (\vec q\,)}
\,
\frac{M_{B_k}}{E_k (-\vec{q}\,)}
}
+ \left. \frac{n(\vec{q}_{\rm lab})}{\sqrt{s} + \omega_k
(\vec{q}\,) -
E_k
(-\vec{q}\,) - i \epsilon} \right\} \ ,
\label{eq:gpauli}
\end{eqnarray}
where $n(\vec{q}_{\rm lab})=\theta\left(p_F - \left|\vec{q}_{\rm lab}\right|\right)$ with $\vec{q}_{\rm lab}$ being
the nucleon momentum in the lab frame, $(P_0,\vec{P}\,)$ is the total four-momentum and $\rho = \frac{2p_F^2}{3 \pi^2}$ is the density of symmetric nuclear matter.

We also consider the in-medium effects on the properties of the baryons $(N, \Lambda, \Sigma)$ by adding a potential term to their free-space energy. As motivated in Ref.~\cite{ramos:2000npa}, we consider the following density dependence of the potentials:
\begin{align}
    V_N = & -70\frac{\rho}{\rho_0}~\text{MeV},\\ 
    V_{\Lambda} = & -58\frac{\rho}{\rho_0} + 31 \left(\frac{\rho}{\rho_0}\right)^2~\text{MeV}, \\ \label{eq:V_baryon}
    V_{\Sigma} = &~ 30 \frac{\rho}{\rho_0}~\text{MeV}. 
\end{align}
where $\rho_0 = 0.17$~fm$^{-3}$ is the saturation density of nuclear matter. 
It is to be noted that our model is sensitive only to the total $\Lambda$ potential in the nuclear medium, which corresponds to $V_{\Lambda}(\rho_0) = -26.4$~MeV, a value well constrained by hypernuclear data. Different distributions of the $\Lambda N$ and $\Lambda NN$ components of the potential, as recently proposed \cite{friedman:npa2023, friedman:PoS2024, weise:epj2020}, will not influence our results. 

Finally, we consider the in-medium self-energy of the
${\bar K}$ and $\pi$ mesons, $\Pi_M(q^0,\vec{q},\rho)$ ($M={\bar K}, \pi$),  as they are the ones interacting more strongly with the nucleons of the Fermi sea. We treat the $\eta$ and $K$ mesons as free propagating particles. 
For the pion self-energy we take that of Ref.~\cite{Ramos:1994xy}, consisting of a small s-wave term plus a p-wave component built from the coupling of the pion to particle-hole,
$\Delta$-hole and two-particle-two-hole excitations. The self-energy of the ${\bar K}$ meson contains an s-wave component, built up from the s-wave ${K^-p}$ and ${K^-n}$ amplitudes as:
\begin{equation}
\Pi^s_{K^-}(q^0,{\vec q},\rho)=2\int \frac{d^3p}{(2\pi)^3}
n(\vec{p}\,) \left[ t_{K^- p \rightarrow K^- p}(P^0,\vec{P},\rho) +
t_{K^- n \rightarrow K^- n}(P^0,\vec{P},\rho) \right] \ ,
\label{eq:selfka}
\end{equation}
where $(q^0,\vec{q}\,)$ is the four-momentum of the
$K^-$ in the laboratory frame, while $P^0=q^0+E(\vec{p}\,)$ and $\vec{P}=\vec{q}+\vec{p}$ are the
total energy and momentum of the $K^-N$ pair in this frame. The $K^-$ self-energy of Eq.~\eqref{eq:selfka} requires to be evaluated self-consistently, since the in-medium amplitudes 
$t_{K^- N \rightarrow K^- N}$ ($N=p,n$), which build up the self-energy $\Pi^s_{K^-}$, depend on this self-energy themselves, as we will see.
The self-energy also contains
a p-wave component constructed from the coupling of the $K^-$ meson to (nucleon)hole-hyperon excitations. The details can be found in Ref.~\cite{ramos:2000npa}. 
 
In terms of the meson self-energy, the in-medium dressed propagator reads ($M=\bar{K},\pi$):
\begin{equation}
D_M(q^0,\vec{q},\rho) = \frac{1}{(q^0)^2-{\vec q\,}^2 - m_M^2 -
\Pi_M(q^0,\vec{q},\rho)} \ ,
\end{equation}
while the corresponding Lehmann representation is:
\begin{eqnarray}
D_M(q^0,\vec{q},\rho) &=& \int_0^\infty d\omega
\frac{S_M(\omega,\vec{q},\rho)}{q^0-\omega + i\epsilon} \,  -
\int_0^\infty d\omega \frac{
S_{\bar{M}}(\omega,\vec{q},\rho)}{q^0 + \omega - i\epsilon}  \ ,
\label{eq:dressed}
\end{eqnarray}
where the spectral
density:
\begin{equation}
S_{M(\bar{M})}(\omega,{\vec
q},\rho)= -\frac{1}{\pi} {\rm Im}\, D_{M(\bar{M})}(\omega,{\vec q},\rho) =
-\frac{1}{\pi}\frac{{\rm Im} \Pi_{M(\bar{M})}(\omega,\vec{q},\rho)}
{\mid \omega^2-\vec{q}\,^2-m_M^2-
\Pi_{M(\bar{M})}(\omega,\vec{q},\rho) \mid^2} \ ,
\end{equation}
has been introduced. With the dressed propagator, the loop function becomes:
\begin{eqnarray}
G_k^\Lambda(P^0,\vec{P},\rho)&= &
\int_{\mid {\vec q} \mid < \Lambda} \frac{d^3 q}{(2 \pi)^3}
\frac{m_{B_k}}{E_{B_k} (-\vec{q}\,)} \nonumber \\
&\times & \left[\int_0^\infty d\omega 
 S_{M_k}(\omega,{\vec q},\rho) 
\frac{1-n(\vec{q}_{\rm lab})}{\sqrt{s}- \omega
- E_{B_k} (-\vec{q}\,)
+ {\rm i} \epsilon} \right. \nonumber \\
&+& \left. \int_0^\infty d\omega  S_{\bar{M}_k}(\omega,{\vec q},\rho)
\frac{n(\vec{q}_{\rm lab})}
{\sqrt{s} + \omega - E_{B_k}(-\vec{q}\,) } \right] \ ,
\label{eq:gmed}
\end{eqnarray}
where $n(\vec{q}_{\rm lab})=0$ for channels
involving hyperons.

The in-medium amplitudes are then obtained by solving the Bethe-Salpeter equation [Eq.~(\ref{eq:BS})] replacing the free meson-baryon propagator loop function by the in-medium one for channels involving ${\bar K}$ or $\pi$ mesons. Note that, since the free scattering amplitudes of the BCN model have been obtained employing a dimensional regularization scheme, a proper inclusion of the in-medium effects requires defining the in-medium loop as:
\begin{equation}
    G_k(P^0,\vec{P},\rho)=G^{DR}_k(\sqrt{s}) + G_k^{\Lambda}(P^0,\vec{P},\rho) - G_k^{\Lambda}(\sqrt{s})  \ ,
\end{equation}
 where $G^{DR}_k(\sqrt{s})$ is given by Eq.~(\ref{eq:GmatrixDR}), $G_k^{\Lambda}(\sqrt{s})$ by Eq.~(\ref{eq:gprop}) and $G_k^{\Lambda}(P^0,\vec{P},\rho)$ by Eq.~(\ref{eq:gpauli}) or by Eq.~(\ref{eq:gmed}), depending on the considered approximation  for in-medium effects.

\begin{figure}[t!]
    \centering
    \includegraphics[width=0.9\textwidth]{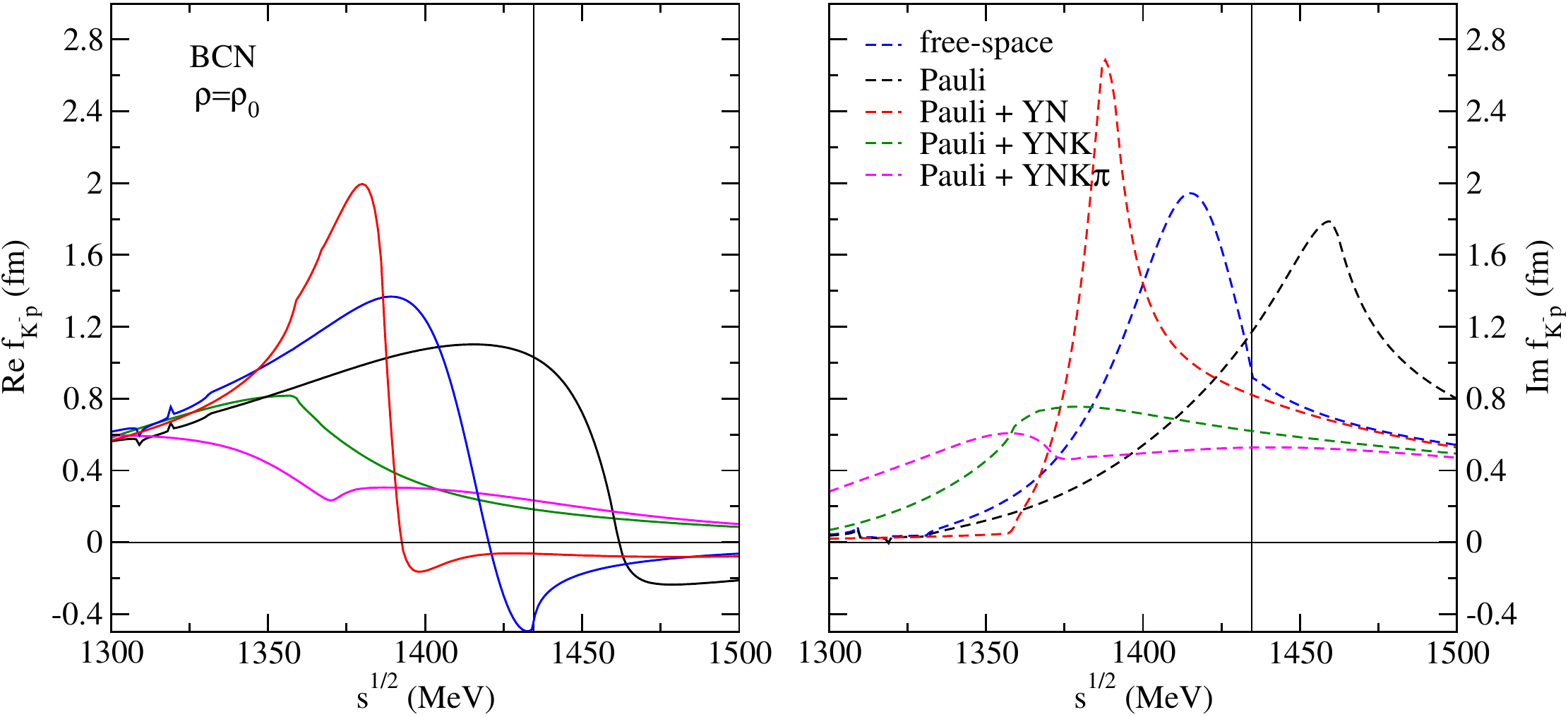}
    \caption{Real (left) and imaginary (right) parts of the $K^-p\rightarrow K^-p$ amplitude, $f_{K^-p}$, with Pauli blocking (black), Pauli+YN SE (red), Pauli + YNK SE (green) and Pauli + YNK$\pi$ SE (magenta) calculated at $\rho=0.17$~fm$^{-3}$. The corresponding free-space amplitude (blue) is shown for comparison.}
    \label{fig:1}
\end{figure}

\begin{figure}[h!]
    \centering
    \includegraphics[width=0.9\textwidth]{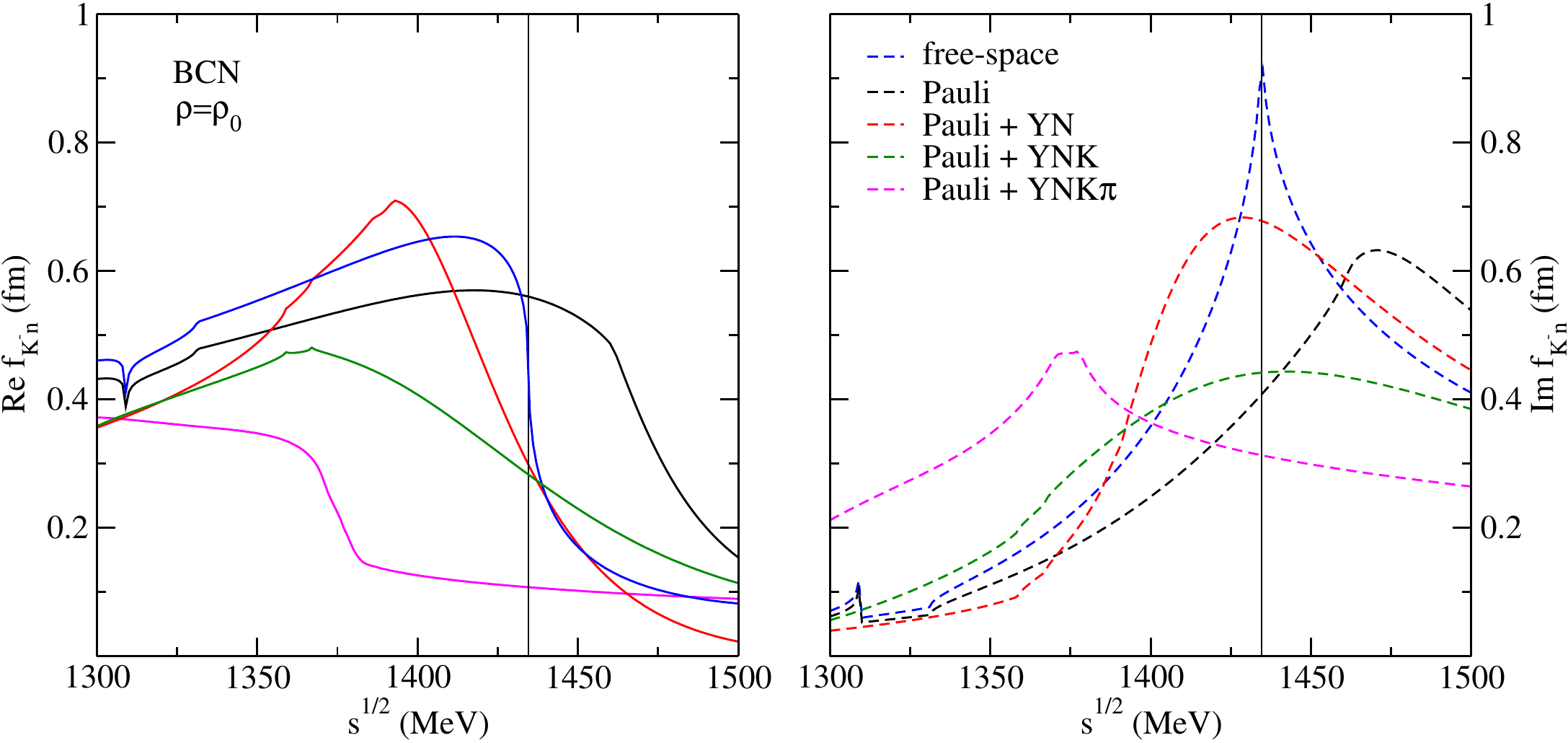}
    \caption{Real (left) and imaginary (right) parts of the $K^-n\rightarrow K^-n$ amplitude, $f_{K^-n}$, with Pauli blocking (black), Pauli+YN SE (red), Pauli + YNK SE (green) and Pauli + YNK$\pi$ SE (magenta) calculated at $\rho=0.17$~fm$^{-3}$. The corresponding free-space amplitude (blue) is shown for comparison.}
    \label{fig:2}
\end{figure}

In Figs.~\ref{fig:1} and \ref{fig:2} we display the $K^-p$ and $K^-n$ amplitudes, respectively, in symmetric nuclear matter at saturation density, for different in-medium approximations: considering only Pauli blocking (Pauli), including also hyperon and nucleon self-energies (Pauli+YN), dressing of $K^-$ in the medium (Pauli+YNK), and finally, dressing of pion in the medium (Pauli+YNK$\pi$).  The left (right) panels show the real (imaginary) part of the amplitudes. When Pauli-blocking effects are incorporated (black lines) the amplitudes are displaced to larger invariant mass energies compared to the free-space ones (blue lines). This is essentially tied to the loss of phase space for the intermediate ${\bar K}N$ states, since the excited nucleon is required to be above the Fermi level. The additional incorporation of the baryon potentials (red lines) move the structures of the amplitudes towards lower energies, essentially by 70 MeV, which is the potential felt by nucleons at $\rho=\rho_0$. The self-consistent incorporation of the antikaon self-energy (green lines) smooths the amplitudes substantially due to the fact that the antikaon spectral function spreads over a wider energy range compared to the delta-function behavior of the previous cases. Incorporating the pion self-energy effects (magenta lines) produces an additional smoothening and, especially, a substantial increase of the imaginary part of the amplitudes below 1350 MeV. This is due to the coupling of the pion to 1p1h and 2p2h states, which effectively lowers the energy of the available states with $\bar{K}N$ quantum numbers down to the hyperon mass. The kink that appears around 1370 MeV, which signals the opening of the $\pi\Sigma$ channel, is enhanced precisely because of the presence of 1p1h$\Sigma$ and 2p2h$\Sigma$ excitations that are already open at lower energies.

\begin{figure}[h!]
    \centering
    \includegraphics[width=0.7\textwidth]{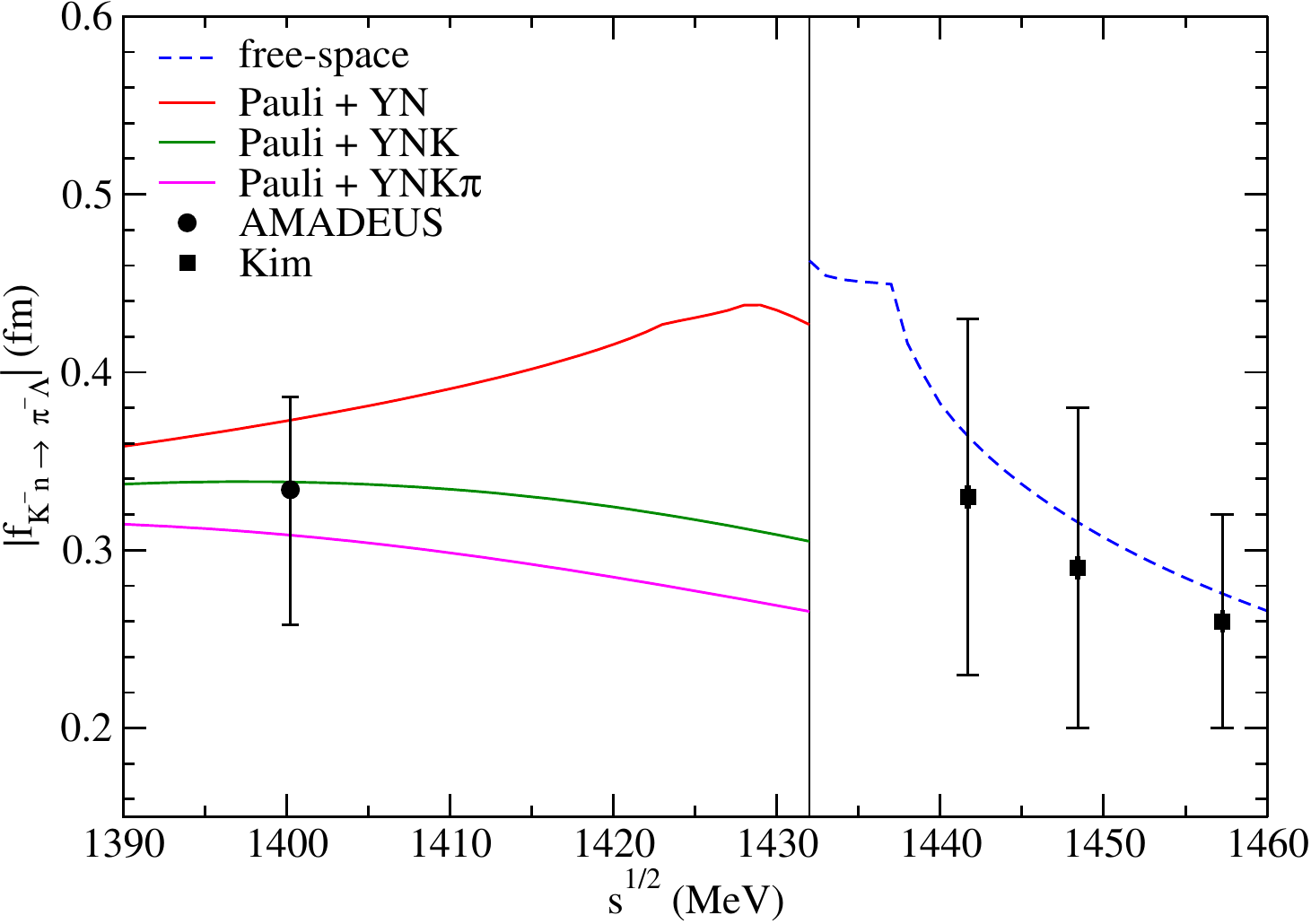}
    \caption{Absolute value of the $K^-n\rightarrow \pi^- \Lambda$ amplitude in the Pauli+YN SE (red), Pauli + YNK SE (green) and Pauli + YNK$\pi$ SE (magenta) models, evaluated at $\rho = 0.3\rho_0$ as a function of the centre-of-mass energy $\sqrt{s}$ compared with the measurement by the AMADEUS collaboration \cite{amadeus2018}. For comparison there is also the corresponding free-space amplitude (blue) above threshold (vertical line) compared with the amplitude extracted from cross section data~\cite{kim_exp}.}
    \label{fig:3}
\end{figure}

Finally, we present in Fig.~\ref{fig:3} two observables related to the isospin $I=1$ transition amplitude $K^- n \to \pi^- \Lambda$. On the right-hand side of the plot, the modulus of the free-space amplitude is compared to that extracted from the inelastic $K^- p \to \pi^0 \Lambda$ cross section at $K^-$ laboratory momenta of 120 MeV, 160 MeV and 200 MeV \cite{kim_exp}. On the left-hand side, our in-medium amplitudes at $\rho=0.3 \rho_0$ are compared with the measurement by the AMADEUS collaboration \cite{amadeus2018} on the absorption at rest of $K^-$ by $^4$He, which delivered the modulus of the $K^- n \to \pi^- \Lambda$ amplitude about 33 MeV below the $K^- n$ threshold, a shift that accounts for the 21 MeV binding of the neutrons and the recoil energy of the $\pi^- \Lambda$ pair with respect to the residual nucleus $^3$He. The size of the error bars does not permit discriminating between the in-medium models, although it can be seen that the consideration of meson self-energies brings the results (green and magenta lines) closer to the central value of the experiment.

\subsection{$K^-N$ and $K^-NN$ absorption model including hadron self-energies}
\label{KNNmodel}

In Ref.~\cite{hrtankova:2020prc}, we developed a microscopic model for the $K^-N$ and $K^-NN$ absorption in nuclear matter. We employed the chiral $K^-N$ amplitudes derived within the BCN \cite{feijoo:2019prc} and Prague \cite{pnlo} models. We took into account the Pauli blocking effect in the amplitudes, but they lacked the proper inclusion of the hyperon, kaon, and pion self-energies, which we address in the present study.

\begin{figure}[t!]
    \centering
    \includegraphics[width=0.2\textwidth]{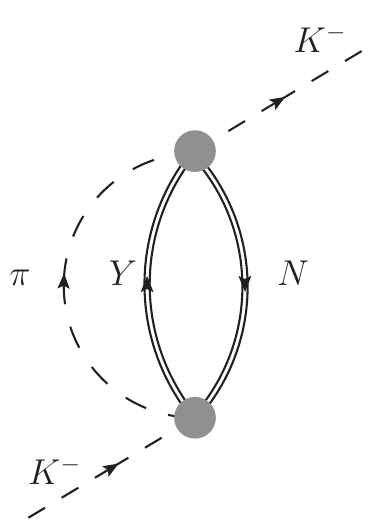}
    \caption{Feynman diagram corresponding to the $K^-N\rightarrow \pi Y$ ($Y=\Sigma, \Lambda$) absorption channel in nuclear matter with dressed hyperon and nucleon.}
    \label{fig:4}
\end{figure}

Similarly as in our previous work \cite{hrtankova:2020prc}, the imaginary part of the $K^-N$ potential derived from the Feynman diagram in Fig.~\ref{fig:4} is of the form
\begin{equation}\label{eq:imVknpiY+YN_se}
\text{Im}V_{K^-N\rightarrow \pi Y}(p) = \frac{\text{Im} \Pi_{K^-}(p)}{2 E_{K^-}}= -\frac{1}{2 E_{K^-}}\frac{1}{4\pi} \frac{\rho}{2}~|t_{K^-N\rightarrow \pi Y}|^2 \frac{\overline{q}}{\langle E_N \rangle} \frac{m_N m_Y}{E_Y(\overline{q}) + \omega_{\pi}(\overline{q})}~. 
\end{equation}
Here, $\Pi_{K^-}$ denotes the kaon self-energy, $p=(p_0, \vec{p}_{K^-})$ is the kaon four-momentum, $E_{K^-}=m_{K^-}-B_{K^-}$ is the kaon energy with $m_{K^-}$ being the $K^-$ mass and $B_{K^-}$ being the kaon binding energy, $\langle E_N \rangle = \sqrt{m_N^2 + \frac{3}{5} k_F^2} + V_N$ is the average nucleon energy, $m_N$ is the nucleon mass and $k_F$ is the Fermi momentum. The quantity $t_{K^-N\rightarrow \pi Y}$ is the t-matrix derived from the chiral model and $\omega_{\pi}(\bar{q})$ is the pion energy. The hyperon energy is given by $E_Y(\overline{q})=\sqrt{m_Y^{\prime~2} + \overline{q}\,^{2}}+V_Y$ ($Y=\Lambda, \Sigma$), with $m_Y^{\prime\, 2}=m_Y^2 + \frac{3}{5}k_F^2 + p^2_{K^-}$, $m_Y$ being the hyperon mass, and  $\overline{q}$ the on-shell pion momentum stemming from energy conservation,
\begin{equation}
 \overline{q}\,^2 = \frac{\left(s_{KN}-(m_Y^{\prime}+m_{\pi})^2\right)\left(s_{KN}-(m_Y^{\prime}-m_{\pi})^2\right)}{4s_{KN}}~,
\end{equation}
where $s_{KN}=(E_{K^-}+\langle E_N \rangle-V_Y)^2$ and $m_{\pi}$ is the pion mass. Note that the momentum  $\overline{q}$ as well as the hyperon energy $E_Y$ are now modified due to the insertion of the hyperon potential $V_Y$.

\begin{figure}[h!]
    \centering
    \includegraphics[width=0.48\textwidth]{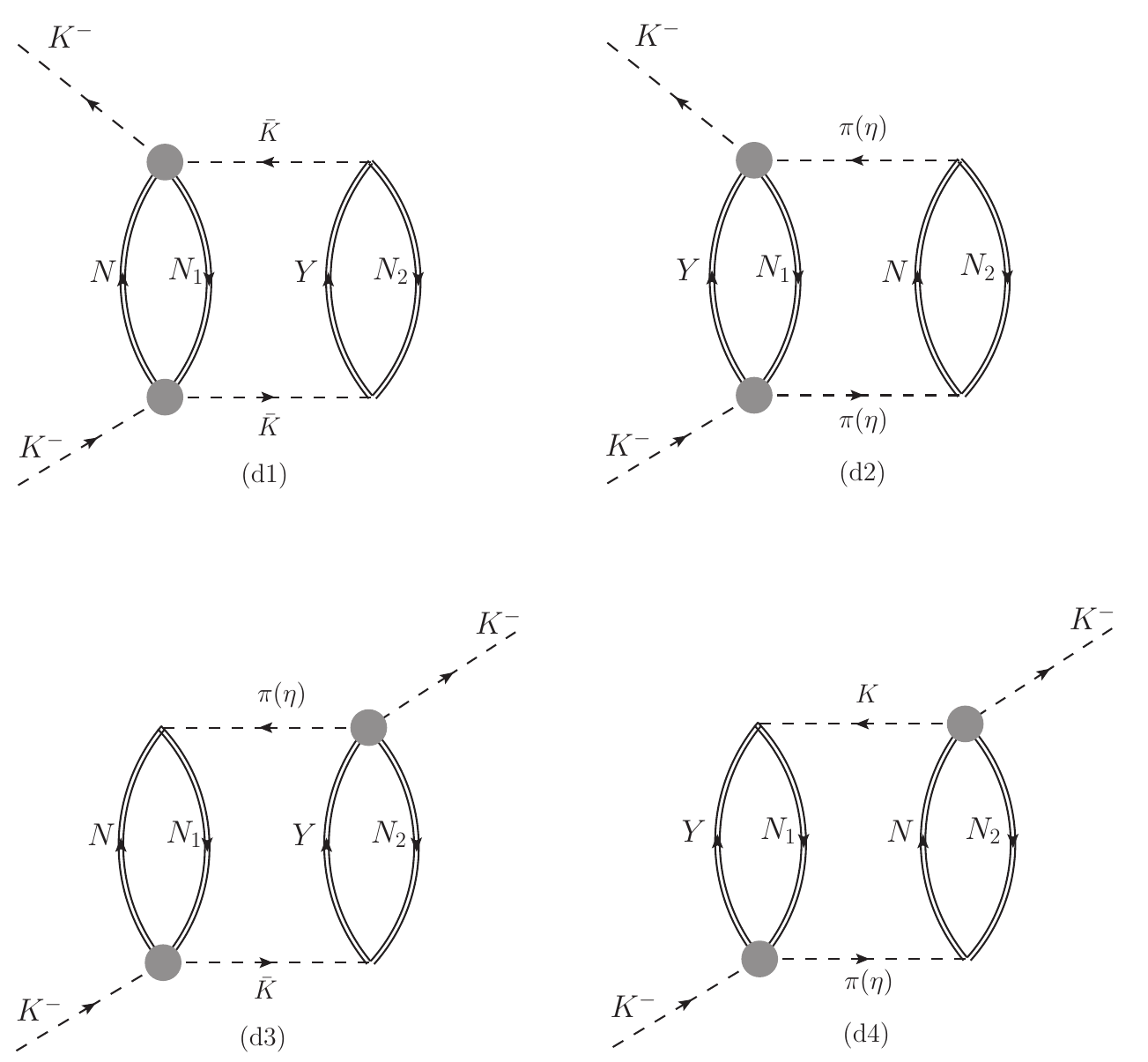} \hspace{10pt}
    \includegraphics[width=0.48\textwidth]{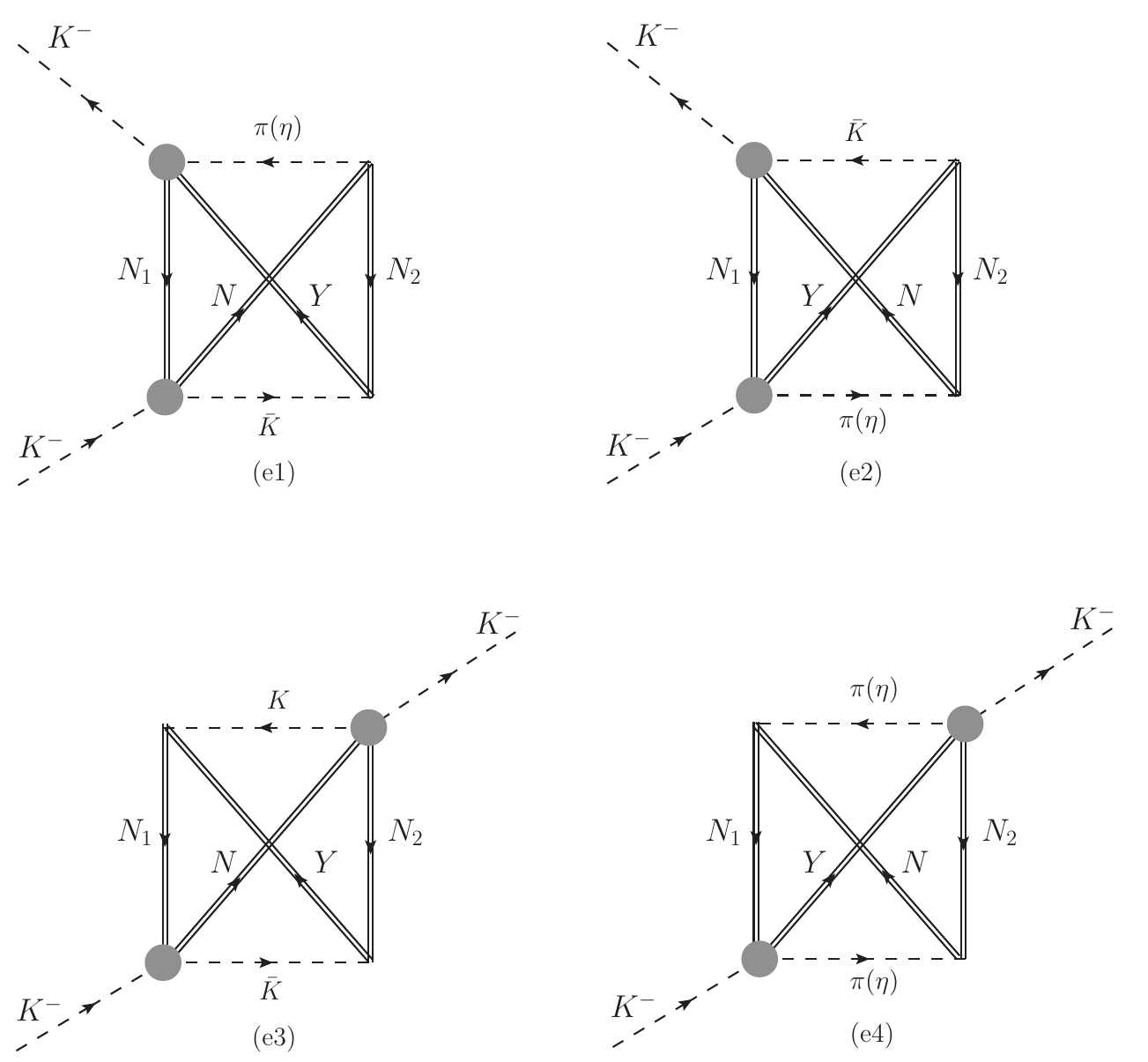}
    \caption{Feynman diagrams corresponding to the $K^-NN\rightarrow N Y$ ($Y=\Sigma, \Lambda$) absorption channels in the nuclear matter with dressed hyperons and nucleons.}
    \label{fig:5}
\end{figure}

In Fig.~\ref{fig:5}, there are Feynman diagrams representing the $K^-$ absorption on 2 nucleons at rest in nuclear medium. The double-lines represent the in-medium baryons  dressed by a corresponding potential. The main contribution to the absorption comes from the diagrams (d1) and (d2) which we further denote as two-fermion-loop (2FL) diagrams. The diagrams in (d3), (d4), (e1) - (e4) result from antisymmetrization of initial $N_1$ and $N_2$ lines as well as $Y$ and $N$ lines in the final state. It is to be noted that the contribution from (d3) and (d4) diagrams is zero due to the null trace over spins. We refer to diagrams (e1) and (e2) as one-fermion-loop of type A (1FLA) and to diagrams (e3) and (e4) as one-fermion-loop of type B (1FLB). The derivation of the expressions corresponding to the diagrams of Fig.~\ref{fig:5} is detailed in Ref.~\cite{hrtankova:2020prc}. For completeness, we include the most relevant results in what follows.
The expressions for Re$V_{K^-NN}$ and Im$V_{K^-NN}$ corresponding to the 2FL diagrams (d1) and (d2) in Fig.~\ref{fig:5} read, respectively,
\begin{align}\label{eq:reV2FL}
 \text{Re}V_{K^-NN}^{\rm 2FL}(p) =  \frac{\text{Re}\Pi_{K^-NN}^{\rm 2FL}(p)}{2E_{K^-}} =&
 \frac{1}{2 E_{K^-}} t_{B_1 x} t^*_{B_1 x}  V_{B_2 N_2 x} V_{B_2 N_2 x} \frac{\rho^2}{4} \nonumber \\
& \times
 \int \frac{\vec{q}^{\,2} dq}{2 \pi^2} \left(\frac{m_N}{\langle E_N \rangle}\right)^2 \frac{m_{B_1}}{E_{B_1}} \frac{m_{B_2}}{E_{B_2}}~ \vec{q}^{\,2}  \omega(\vec{q}\,) F^2_H(\vec{q}\,)~ \nonumber \\[1ex] & \times \frac{1}{E_{K^-}+2\langle E_N \rangle- E_{B_1}(\langle \vec{j} \rangle + \vec{p}_{K^-}-\vec{q}\,)-E_{B_2}(\langle \vec{k} \rangle+\vec{q}\,)}  \notag \\[1ex] &  \times \frac{1}{q_0^2-\vec{q}^{\,2}-m_{x}^2}~ \frac{1}{q_0^2-\vec{q}^{\,2}-m_{x}^2}~.
\end{align}

\begin{align}
 \text{Im}V_{K^-NN}^{\rm 2FL}(p)= \frac{\text{Im} \Pi_{K^-NN}^{\rm 2FL}(p)}{2 E_{K^-}} =& -\frac{1}{2 E_{K^-}}  t_{B_1 x} t^*_{B_1 x} 
 V_{B_2 N_2 x} V_{B_2 N_2 x} \frac{1}{2 \pi} \frac{\rho^2}{4} \nonumber \\
 & \times \frac{m_{B_1} m_{B_2}}{E_{B_1}(\overline{q})+E_{B_2}(\overline{q})} \left(\frac{m_N}{\langle E_N \rangle}\right)^2 \overline{q} \notag \\[1ex] & \times \overline{q}^{\,2} \omega(\overline{q}) F^2_H(\overline{q})~ \frac{1}{q_0^2-\overline{q}^{2}-m_{x}^2}~ \frac{1}{q_0^2-\overline{q}^{2}-m_{x}^2}~.
\label{eq:imV2FL}
\end{align}
Here, $t_{B_1x}$ corresponds to two-body $t$ matrix for the $K^-N \rightarrow B_1x$ channel, where $x$ denotes the intermediate meson exchanged and $B_1$ corresponds to baryon ($N$ or $Y$) attached to the incoming kaon vertex. The Yukawa p-wave type meson-baryon-baryon coupling vertex is denoted as $V_{B_2N_2x}=\alpha \frac{D+F}{2f_{\pi}} + \beta \frac{D-F}{2f_{\pi}}$, with $D+F = 1.26$, $D-F = 0.33$, $f_{\pi}=93$~MeV, and $\alpha, \beta$ being the appropriate SU(3) Clebsch-Gordan coefficients, and $B_2$ denotes the baryon ($Y$ or $N$) emitted at the vertex. The quantities $q_0, \vec{q}, m_x$ are the virtual meson energy, momentum, and mass, respectively, $\langle \vec{j}\rangle = \langle \vec{k} \rangle = \sqrt{\frac{3}{5}}k_F $ are the averaged nucleon momenta. The factor $\omega(\bar{q})$ is the angle average value of the Pauli function $\theta(|\langle \vec{j}\rangle+\vec{p_{K^-}}-\vec{q}|-k_F)$ and $F_H(\bar{q})= \frac{\Lambda^2_C}{\Lambda^2_C+\bar{q}}$ is the form-factor for the Yukawa vertices with a cut-off parameter $\Lambda_C=1200$~MeV. Finally, 
\begin{equation}\label{eq:Ebaryon_2FL}
E_{B_{1(2)}}(\overline{q})=\sqrt{m_{B_{1(2)}}^{\prime~2}+\overline{q}^2} + V_{B_{1(2)}}
\end{equation}
with $m_{B_{1}}^{\prime~2}=m_{B_{1}}^{~2} + \frac{3}{5}k_F^2+p^2_{K^-}$, $m_{B_{2}}^{\prime~2}=m_{B_{2}}^{~2} + \frac{3}{5}k_F^2$, $V_{B_1}=V_N$ or $V_Y$ and $V_{B_2} = V_Y$ or $V_N$, and 
 $\overline{q}$ is center-of-mass momentum of the exchanged meson
 \begin{equation}\label{eq:qbar_meson_2FL}
 \overline{q}^2 = \frac{(s_{K2N}-(m_{B_1}^{\prime}+m_{B_2}^{\prime})^2)(s_{K2N}-(m_{B_1}^{\prime}-m_{B_2}^{\prime})^2)}{4s_{K2N}}~,
\end{equation}
with $s_{K2N}=(E_{K^-}+2\langle E_N \rangle - V_{B_1} - V_{B_2})^2$. The expressions for the imaginary part of the $K^-NN$ potential stemming from the 1FLA diagrams (e1) and (e2) in Fig.~\ref{fig:5} acquire the following form:
\begin{align}\label{eq:ImPiB1B2_c2}
\text{Im}V_{K^-NN}^{\rm 1FLA}(p) = \frac{\text{Im} \Pi_{K^-NN}^{\rm 1FLA}(p)}{2 E_{K^-}} =& - \frac{1}{2 E_{K^-}}\frac{1}{2}  t_{B_1 x_1} t^*_{B_2 x_2} V_{B_2 N_2 x_1} V_{B_1 N_2 x_2} \frac{1}{2 \pi} \frac{\rho^2}{4} \nonumber \\
& \times \overline{q}~ \frac{m_{B_1} m_{B_2}}{E_{B_1}(\overline{q})+E_{B_2}(\overline{q})} \left(\frac{m_N}{\langle E_N \rangle}\right)^2  \overline{q}^{\,2} \omega(\overline{q})~ \notag \\[1ex] & \times \frac{F_H(\overline{q})}{q_0^2-\overline{q}^{\,2}-m_{x_1}^2}~ \frac{F_H(\langle j \rangle+\vec{p}_{K^-}-\overline{q}-\langle k\rangle)}{q_0^{\prime\,2}-\overline{q}^{\,2}-\langle j \rangle^2-\langle k\rangle^2-p^2_{K^-}-m_{x_2}^2}~,
\end{align}
where $q_0^{\prime}=E_{B_1}(\overline{q})-\langle E_N \rangle$. The real part is as follows
\begin{align}\label{eq:RePiB1B2_c2}
 \text{Re}V_{K^-NN}^{\rm 1FLA}(p) =  \frac{\text{Re}\Pi_{K^-NN}^{\rm 1FLA}(p)}{2 E_{K^-}} =& \frac{1}{2 E_{K^-}}\frac{1}{2} t_{B_1 x_1} t^*_{B_2 x_2} V_{B_2 N_2 x_1} V_{B_1 N_2 x_2} \frac{\rho^2}{4} \nonumber \\
 & \times \int \frac{\vec{q}^{\,2} dq}{2 \pi^2} \left(\frac{m_N}{\langle E_N \rangle}\right)^2 \frac{m_{B_1}}{E_{B_1}} \frac{m_{B_2}}{E_{B_2}}~\vec{q}^{\,2}  \omega(\vec{q}\,) \notag \\[1ex] &\times \frac{1}{E_{K^-}+2\langle E_N \rangle- E_{B_1}(\langle \vec{j} \rangle+\vec{p}_{K^-}-\vec{q}\,)-E_{B_2}(\langle \vec{k} \rangle+\vec{q}\,)}\notag \\[1ex] & \times \frac{F_H(\vec{q}\,)}{q_0^2-\vec{q}^{\,2}-m_{x_1}^2}~  \frac{F_H(\langle j \rangle+\vec{p}_{K^-}-\vec{q}-\langle k\rangle)}{q_0^{\prime\,2}-\vec{q}^{\,2}-\langle j\rangle^2-\langle k\rangle^2-p^2_{K^-}-m_{x_2}^2}~,
\end{align}
where variable $E_{B_{1(2)}}$ is defined by Eq.~\eqref{eq:Ebaryon_2FL} and $\bar{q}$ by Eq.~\eqref{eq:qbar_meson_2FL}.

The expressions for the real and imaginary parts of the kaon potential corresponding to the 1FLB-type diagrams (e3) and (e4) in Fig.~\ref{fig:5} are similar to that for the 1FLA-type diagrams (e1) and (e2), with the following changes. The trace over spins yields now a factor $(-2\vec{q}^{~2})$, introducing an overall relative minus sign. There is also a change in the value of $q_0^{\prime}$ which becomes $q_0^{\prime}=\langle E_N \rangle-E_{B_2}(\langle k \rangle+\vec{q}\,)$. Both exchanged mesons in diagrams (e3) and (e4) are now the same, therefore the vertices in Eqs.~(\ref{eq:ImPiB1B2_c2}) and (\ref{eq:RePiB1B2_c2})
are replaced by $t_{B_1 x} t^*_{B_1 x} V_{B_2 N_2 x} V_{B_2 N_2 x}$, with $x$ being either the kaon, pion or eta meson. 
Notice that, due to the spin traces, the 1FL diagrams have acquired an additional factor 1/2 with respect to the 2FL ones. Moreover, the sign of the 1FLA (1FLB) diagrams is the same (opposite than) as that of the 2FL ones. For more details regarding the derivation of $K^-$-two nucleon potentials see Ref.~\cite{hrtankova:2020prc}.

The considered absorption channels for 1N and 2N processes are listed in Table~\ref{tab:channels}. While for 1N absorption each channel corresponds to one Feynman diagram of the type of that in Fig.~\ref{fig:4}, for 2N absorption each channel can proceed via several 2FL, 1FLA, and 1FLB Feynman diagrams, depending  on the possible meson exchanges.
\begin{table}[ht!]
\caption{Channels considered for the $K^-$ single-nucleon (left) and two-nucleon (right) absorption in nuclear matter. }
 \begin{tabular}{cl|cl} \label{tab:channels}
  $K^-N$ & $\rightarrow \pi Y$ & $K^-N_1N_2$  &$\rightarrow YN$ \\ \hline
   $K^-p$& $\rightarrow \pi^0 \Lambda$ & $K^-pp$ & $\rightarrow \Lambda p$ \\
   & $\rightarrow \pi^0 \Sigma^0$ & & $\rightarrow \Sigma^0 p$ \\
   & $\rightarrow \pi^+ \Sigma^-$ & & $\rightarrow \Sigma^+ n$  \\
  & $\rightarrow \pi^- \Sigma^+$ & $K^-pn(np)$ & $\rightarrow \Lambda n$ \\ 
  $K^-n$ & $\rightarrow \pi^- \Lambda$  & & $\rightarrow \Sigma^0 n$ \\ 
  & $\rightarrow \pi^- \Sigma^0$ & & $\rightarrow \Sigma^- p$ \\
  & $\rightarrow \pi^0 \Sigma^-$ & $K^-nn$ & $\rightarrow \Sigma^- n $ \\ 
 \end{tabular}
\end{table}

For the Pauli + YN chiral amplitudes, the total $K^-$ potential is evaluated as a sum of all 1N and 2N absorption channels 
\begin{equation}\label{eq:V_K_YN}
V_{K^-}^{YN} = \sum_{\rm channels} V_{K^-N\rightarrow \pi Y} + V_{K^-NN}^{2FL} + V_{K^-NN}^{1FLA} + V_{K^-NN}^{1FLB}~.
\end{equation}
When the kaon and pion are dressed in the medium, the total $K^-$ potential can no longer be evaluated according to Eq.~\eqref{eq:V_K_YN} since now the $t$ matrix resummation of Pauli+YNK and Pauli+YNK$\pi$ chiral amplitudes contain already 2N absorption processes of the type of diagrams (d1) and (d2) in Fig.~\ref{fig:5}, respectively, (as well as additional contributions from 3N, 4N, ... processes). Therefore, in this case, the total $K^-$ potential in the medium is calculated as follows
\begin{equation}\label{eq:V_K_YNKpi}
    V_{K^-}^{\rm YNK(\pi)} = V_{t\rho} + V_{K^-NN}^{\rm corr}~.
\end{equation}
\begin{figure}[t!]
    \centering
    \includegraphics[width=0.4\textwidth]{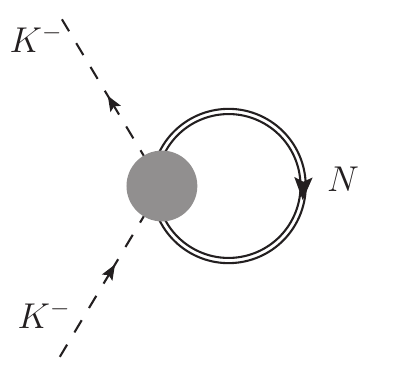}
    \caption{Feynman diagram representing the $K^-$ optical potential in nuclear matter through multiple scattering direct terms.}
    \label{fig:t_rho}
\end{figure}
The $t\rho$ potential in symmetric nuclear matter, corresponding to the Feynman diagram in Fig.~\ref{fig:t_rho}, is of the following form
\begin{equation}\label{eq:V_K_trho}
V_{t\rho} = \frac{1}{2E_{K^-}}\frac{t_{K^-p}+t_{K^-n}}{2} \rho~,
\end{equation}
where $t_{K^-p}$ is the t-matrix corresponding to the $K^-p \rightarrow K^-p$ channel and $t_{K^-n}$ is the t-matrix for the $K^-n \rightarrow K^-n$ channel including kaon and pion self-energies.
The term $V_{K^-NN}^{\rm corr}$ corresponds to the correction for the 2N part not involved in the $t\rho$ potential. In particular, for Pauli+YNK amplitudes the correction is calculated from diagrams (d2) and (e1) - (e4) in Fig~\ref{fig:5}. When we consider also the pion self-energy (Pauli+YNK$\pi$ amplitudes), the correction potential $V_{K^-NN}^{\rm corr}$ is calculated from diagram (d2), considering only $\eta \eta$ and $\eta \pi$ exchange, and diagrams (e1) - (e4) in Fig.~\ref{fig:5}.
The t-matrices (amplitudes) depend on the center-of-mass energy $\sqrt{s}$. For interaction of a $K^-$ with a nucleon in nuclear matter the expression for $\sqrt{s}$ reads
\begin{equation}
\sqrt{s} = \sqrt{(E_{K^-}+ \langle E_N\rangle)^2 - \langle k \rangle^2 - p_{K^-}^2},
\end{equation}
where $E_{K^-}$ is the kaon energy, $\langle E_N\rangle$ is the average nucleon energy, $\langle k \rangle = \sqrt{\frac{3}{5}}k_F$ is the average nucleon momentum and $p_{K^-}$ is the kaon momentum (we average over angles, i.e., $(\langle k \rangle + p_{K^-})^2 \rightarrow \langle k \rangle^2 + p_{K^-}^2$).

\subsection{Kaonic atoms}
\label{atoms}

The above model for the $K^-$ potential was applied in calculations of energy shifts and widths in kaonic atoms. To obtain the $K^-$ binding energy $B_{K^-}$ and width $\Gamma_{K^-}$ we solve the Klein-Gordon equation
\begin{equation}\label{KG}
 \left[ \vec{\nabla}^2  + \tilde{\omega}_{K^-}^2 -m_{K^-}^2 -\Pi_{K^-}(\omega_{K^-},\rho) \right]\phi_{K^-} = 0~,
\end{equation}
where $\tilde{\omega}_{K^-} = m_{K^-} - B_{K^-} -{\rm i}\Gamma_{K^-}/2 -V_C= \omega_{K^-} - V_C$, $\omega_{K^-}$ stands for a complex kaon energy, $V_C$ is 
the Coulomb potential introduced via the minimal substitution~\cite{kkwPRL90}, and  $\rho$ is the nuclear density distribution. The $K^-$ self-energy $\Pi_{K^-}(\omega_{K^-}, \rho)$ is constructed following Eq.~\eqref{eq:V_K_YN} for the Pauli + YN model and Eq.~\eqref{eq:V_K_YNKpi} for the Pauli + YNK and Pauli + YNK$\pi$ amplitude models, by replacing the kaon energy $E_{K^-}$ by the $K^-$-nucleus reduced mass $\mu_{K^-}$, $\Pi_{K^-}^{\rm YNK(\pi)} = 2 {\mu}_{K^-}V_{K^-}^{\rm YNK(\pi)}$.
The $t\rho$ potential is now of the form
\begin{equation}\label{V_trho}
2\mu_{K^-}V_{t\rho}=-4\pi \left(1+\frac{A-1}{A}\frac{\mu_{K^-}}{m_N} \right)\left(F_0(\sqrt{s}, \rho)\frac{1}{2}\rho_p + F_1(\sqrt{s}, \rho)\left(\frac{1}{2}\rho_p+\rho_n\right)\right)~,
\end{equation}
where $F_0(\sqrt{s}, \rho) = -\frac{1}{4\pi}\frac{m_N}{\sqrt{s}}\left(2t_{K^-p} - t_{K^-n}\right)$ and $F_1(\sqrt{s}, \rho) = -\frac{1}{4\pi}\frac{m_N}{\sqrt{s}} t_{K^-n}$ are the isospin 0 and 1 $s$-wave in-medium amplitudes in the $K^-N\rightarrow K^-N$ channel, respectively, depending on the center-of-mass (c.m.) energy $\sqrt{s}$ and medium density $\rho$. Symbols $\rho_p$ and $\rho_n$ denote proton and neutron density distributions, respectively, calculated within the relativistic mean field model TM2 for light and medium mass nuclei ($A<40$) and TM1 for heavy nuclei ($A\geq40$)~\cite{Toki}.
The c.m. energy $\sqrt{s}$ is defined by the Mandelstam variable
\begin{equation}\label{eq:s}
 s=(E_N+E_{K^-})^2-(\vec{p}_N+\vec{p}_{K^-})^2~,
\end{equation}
where $E_N=m_N-B_N$, $E_{K^-}=m_{K^-}-B_{K^-}$ and $\vec{p}_{N(K^-)}$ is the nucleon (kaon) momentum. The momentum dependent term $(\vec{p}_N+\vec{p}_{K^-})^2 
\neq 0$ in the $K^-$-nucleus laboratory frame and generates additional substantial downward energy 
shift \cite{cfggmPLB}. 
The $K^-N$ amplitudes can then be expressed as a function of energy $ \sqrt{s} = E_{\rm th} + \delta \sqrt{s}$ where $E_{\rm th}=m_N + m_{K^-}$. For the relative energy $\delta \sqrt{s}$ we used the expression derived in Refs.~\cite{cfggmPLB,fgNPA2017,hmPRC2017, ofmPRC2022}:
\begin{equation} \label{Eq.:deltaEsLDL}
 \delta \sqrt{s}=  -B_N\frac{\rho}{\bar{\rho}}\, - \beta_N\! \left[B_{K^-}\frac{\rho}{\rho_{\rm max}} + T_N\left(\frac{\rho}{\bar{\rho}}\right)^{2/3}\!\!\!\! +V_C\left(\frac{\rho}{\rho_{\rm max}}\right)^{1/3}\right] + \beta_{K^-} {\rm Re}V_{K^-}(r)~,
\end{equation}
where $\beta_{N(K^-)}={m_{N(K^-)}}/(m_N+m_{K^-})$, $B_N=8.5$~MeV is the average binding energy per nucleon, $T_N = 23$~MeV is the average nucleon kinetic energy, $\rho_{\rm max}$ and $\bar{\rho}$ are the maximal and average value of the nuclear density, respectively. Since $\delta \sqrt{s}$ [ and thus $f_{K^-N}(\sqrt{s})$] depends on Re$V_{K^-}$ and $B_{K^-}$ which by themselves depend on $\sqrt{s}$, $f_{K^-N}(\sqrt{s})$ has to be determined self-consistently by iterations.

\section{Results}
\label{results}

\begin{figure}[t]
    \centering
    \includegraphics[width=0.9\textwidth]{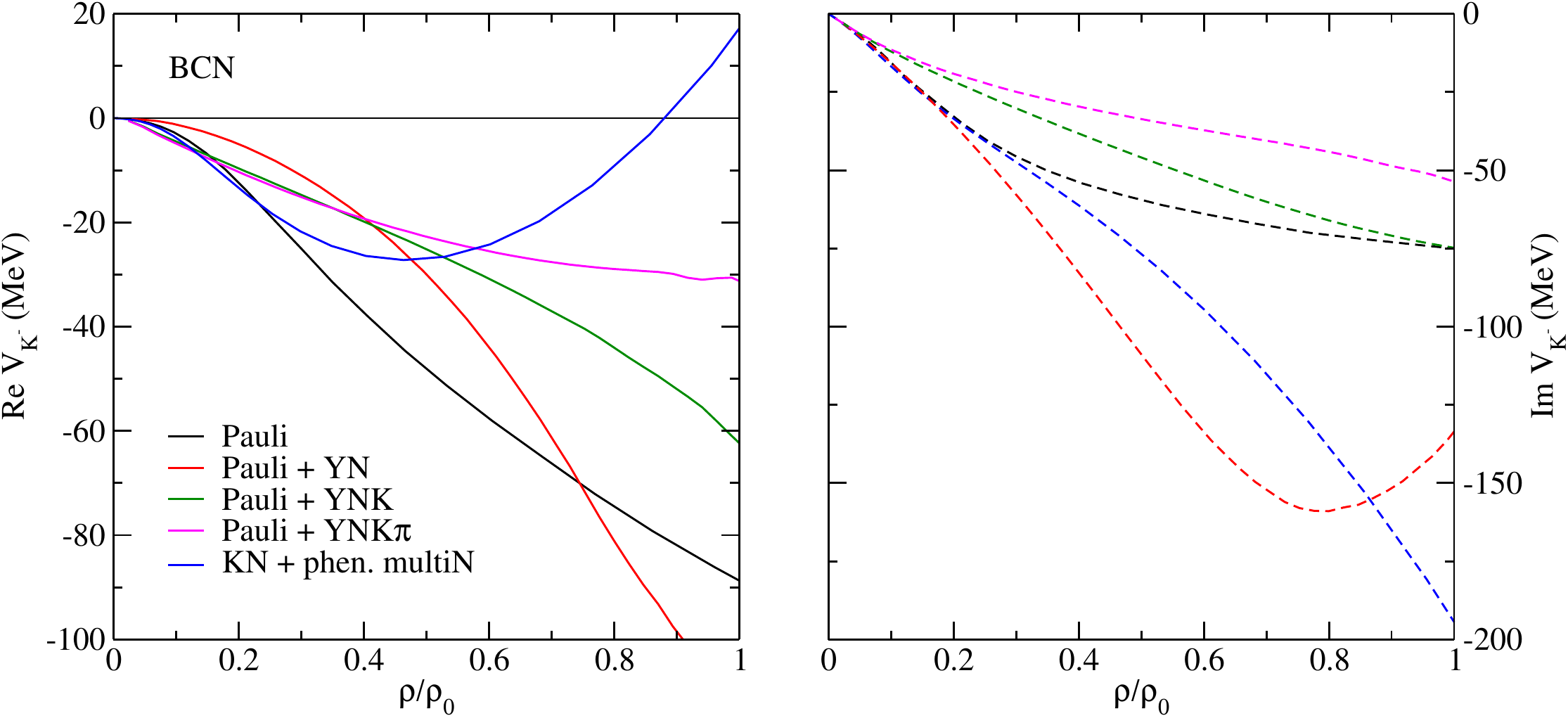}
    \caption{Real (left) and imaginary (right) parts of the total $K^-$ potential calculated with Pauli (black), Pauli +YN SE (red), Pauli + YNK SE (green), and Pauli + YNK$\pi$ (magenta) BCN amplitudes for $B_{K^-}=0$~MeV and $p_{K^-}=0$~MeV/c. For comparison we present the best fit $K^-N$+phen. multiN potential (denoted by 'KN + phen. multiN', blue) based on  BCN amplitudes. }
    \label{fig:6}
\end{figure}

First, we performed calculations of the $K^-$ potential in symmetric nuclear matter. In Fig.~\ref{fig:6}, we present the real (left) and imaginary (right) parts of the total $K^-$ optical potential as a function of the relative nuclear matter density $\rho/\rho_0$. Calculations were performed with the in-medium BCN amplitudes including Pauli blocking (black), Pauli+YN SE (red), Pauli+YNK SE (green) and Pauli+YNK$\pi$ SE (magenta). For comparison, we present also the $K^-N$+phen. multiN potential fitted to reproduce kaonic atom data and single-nucleon absorption fraction from bubble chamber experiments \cite{vander:1977nc, davis:1968nc, moulder:1971npb}. The $K^-N$ part of the $K^-N$+phen. multiN potential is based on the free-space BCN amplitudes modified by the WRW procedure \cite{wrw} to account for Pauli correlation in the medium. The phenomenological multinucleon part
\begin{equation}
    2\mu_{K^-}V_{K^- \rm multiN}^{\rm phen}=-4 \pi B \left(\frac{\rho}{\rho_0}\right)^{\alpha} \rho
\end{equation}
is evaluated for parameters $B=(-1.3, 1.9)$~fm and $\alpha=1$ fitted to kaonic atom data. The fit to data for the $K^-N+$phen. multiN potential yielded $\chi^2(65) = 112.3$. 
It is to be noted that kaonic atom data can reliably determine the shape of Re$V_{K^-}$ up to $0.3\rho_0$ and Im$V_{K^-}$ up to $0.5\rho_0$; beyond this limit the shape of the multinucleon potential is a mere extrapolation of the formula.
The chiral model including only the Pauli blocking yields the real part of the optical potential fairly attractive, reaching $\approx -90$~MeV at $\rho_0$ and the imaginary part is $\approx 75$~MeV deep at $\rho_0$. Including hyperon and nucleon self-energies in the model (Pauli + YN) causes the real part of the potential to be less attractive for $\rho \leq 0.7 \rho_0$ than the Pauli potential, but reaching an overall depth of nearly $120$~MeV at $\rho_0$. The imaginary part of the Pauli + YN potential is more absorptive than the Pauli potential for $\rho > 0.2 \rho_0$. Dressing the kaon causes that both parts of the potential to become shallower due to the weaker chiral amplitudes following the trend in Figs.~\ref{fig:1} and \ref{fig:2}. Finally, dressing also the pion further decreases the depth of the $K^-$ potential in the medium, leading to Re$V_{K^-}(\rho_0)\approx-30$~MeV and Im$V_{K^-}(\rho_0)\approx-50$~MeV.

\begin{figure}
    \centering
    \includegraphics[width=0.8\linewidth]{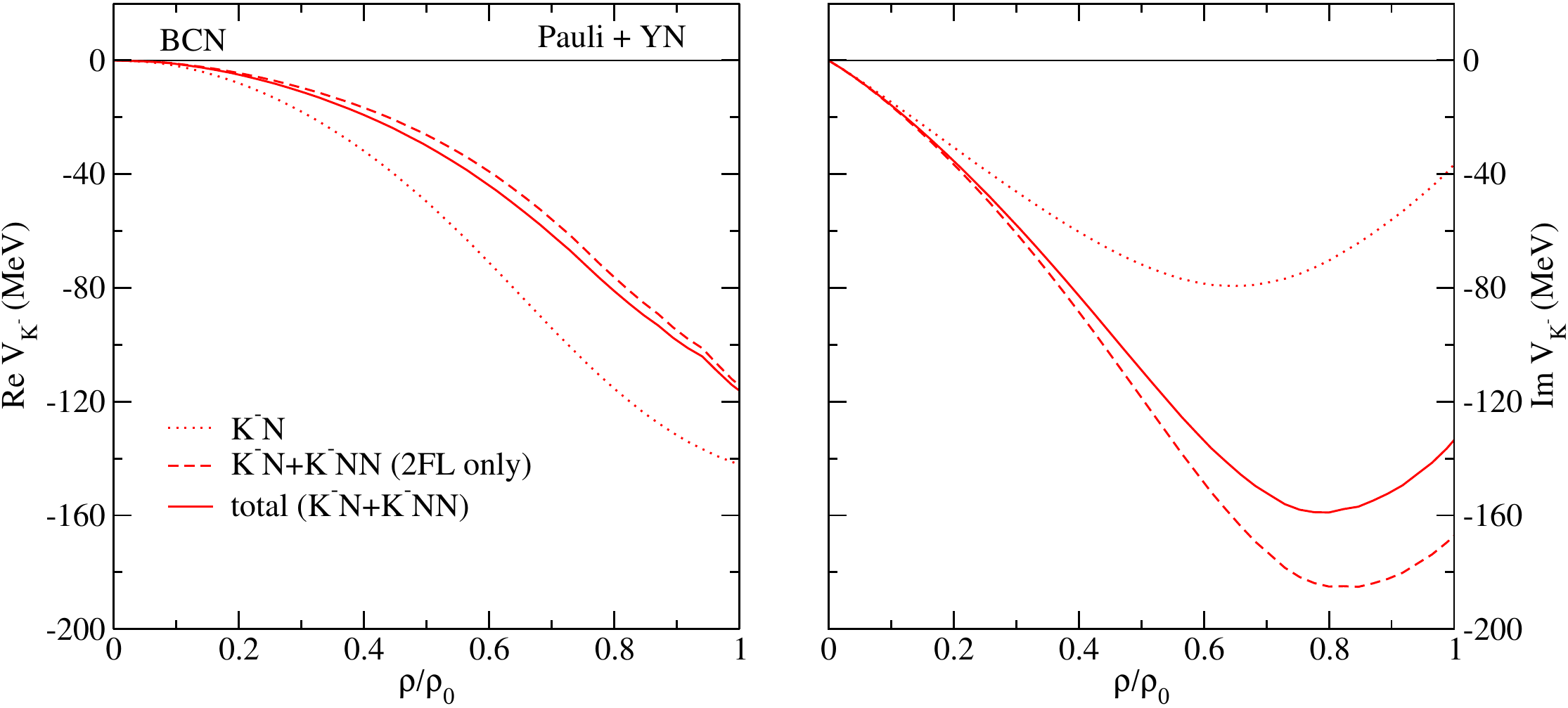}
     \includegraphics[width=0.8\linewidth]{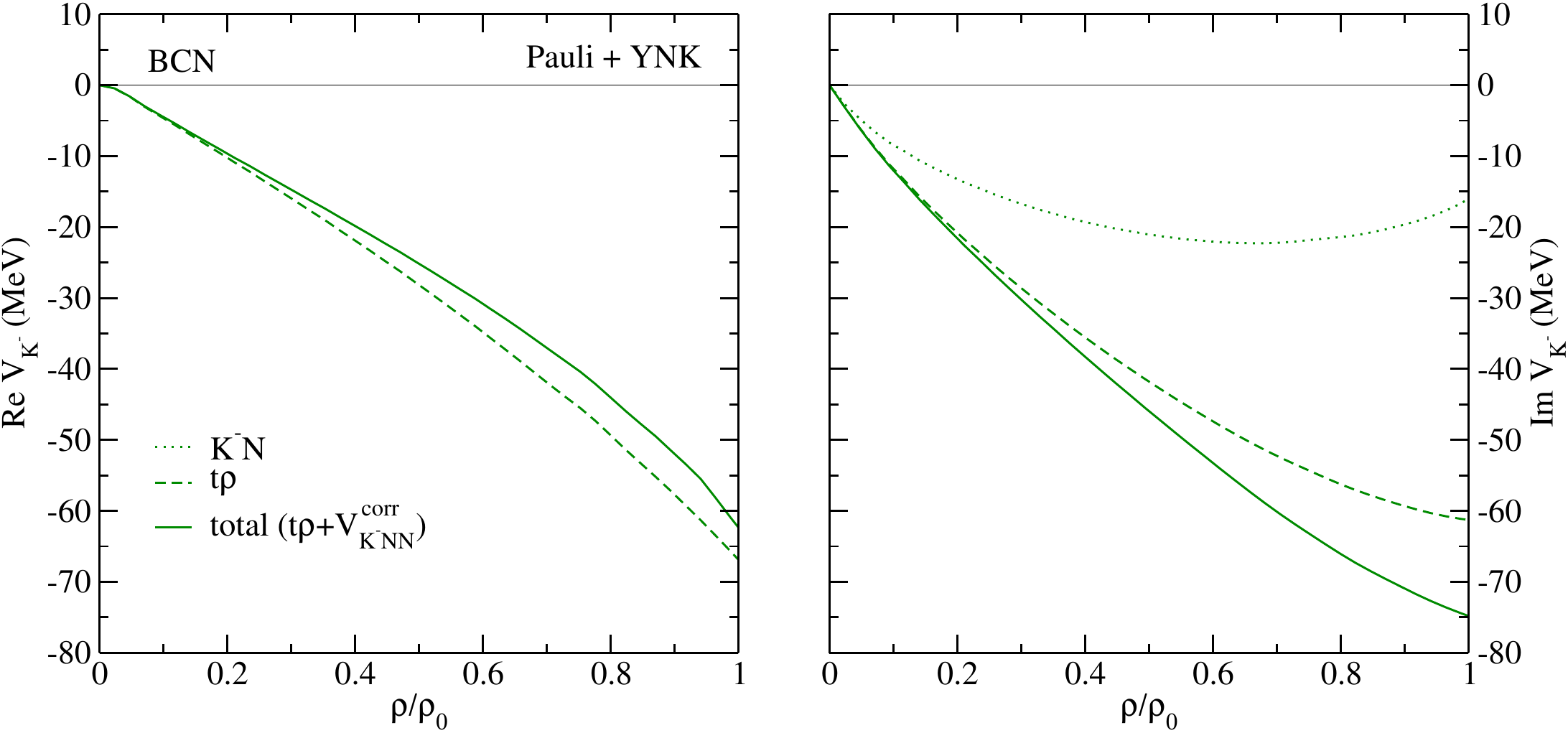}
      \includegraphics[width=0.8\linewidth]{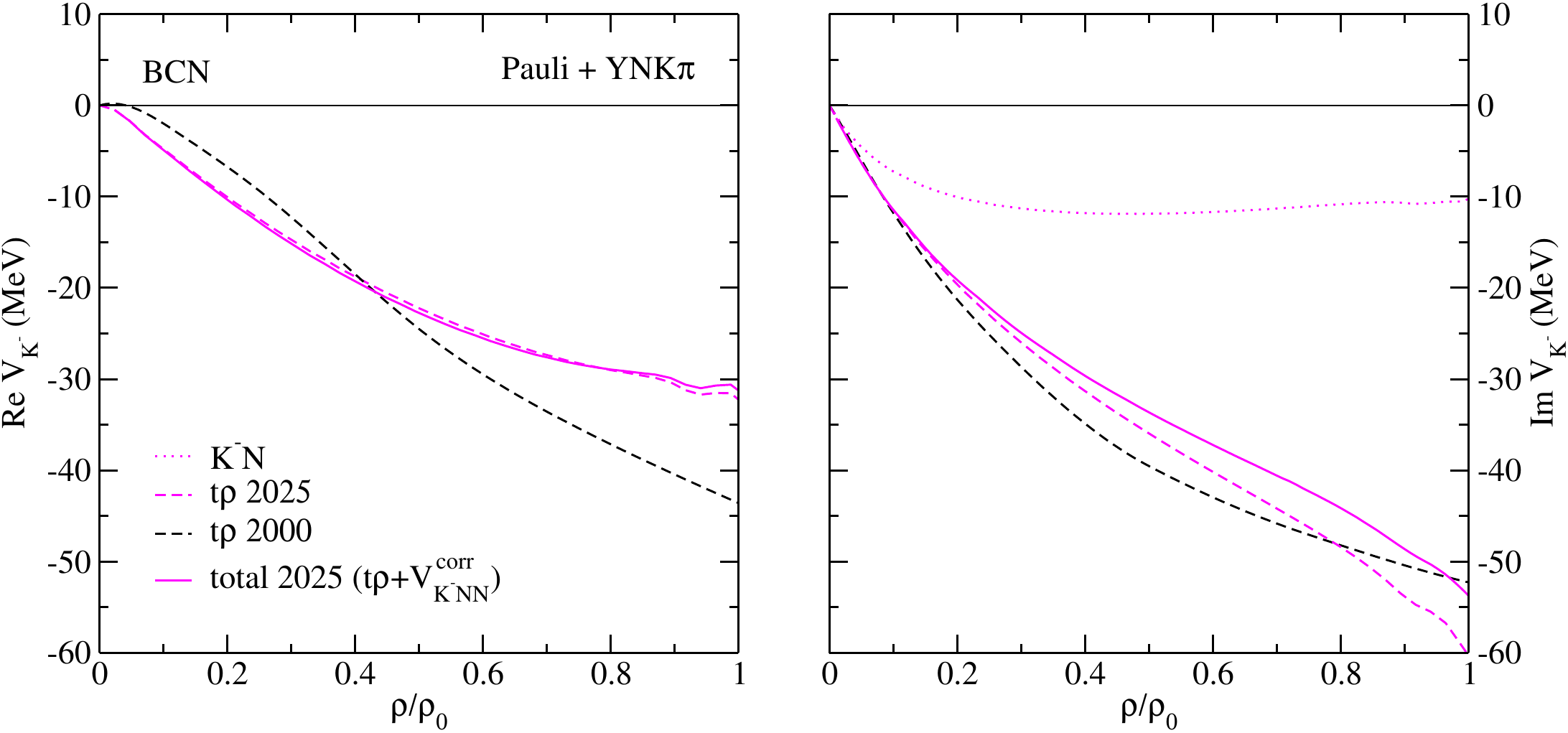}
    \caption{Respective contributions to the total real (left) and imaginary (right) parts of the $K^-$ potential (solid) for considered amplitude models as functions of relative density $\rho/\rho_0$. Top panel: $K^-N$ potential (dotted) and $K^-N+K^-NN$(2FL diagrams only) potential (dashed) obtained with the Pauli + YN SE amplitudes. Middle panel: $K^-N$ potential (dotted) and $t\rho$ part (dashed) of the total $K^-$ potential (solid) calculated within the Pauli+YNK SE model. Bottom panel: the same as for middle panel but for the Pauli+YNK$\pi$ SE amplitude model. For comparison, there is the $K^-$ optical potential including kaon and $\pi$ SE from Ref.~\cite{ramos:2000npa} denoted as '$t\rho~2000$' (black dashed).}
    \label{fig:7}
\end{figure}

In Fig.~\ref{fig:7}, we present respective contributions from the $K^-$-single nucleon, $K^-$-two nucleon and $K^-$-multinucleon potentials in the considered amplitude models as functions of relative density $\rho/\rho_0$. In the top panels of the figure, there are contributions from the $K^-N$ potential (dotted) stemming from diagram in Fig.~\ref{fig:4}, the $K^-N+K^-NN$(2FL only) potential (dashed) including contributions to 2N potential only from the 2FL diagrams (d1) and (d2) in Fig~\ref{fig:5}, and the total $K^-$ potential ($K^-N+K^-NN$, solid) calculated with the Pauli + YN amplitude model. The difference between the dashed and solid lines is due to the contribution from the 1FL exchange-type diagrams (e1) - (e4) of Fig.~\ref{fig:5}, the imaginary part of which is positive. The total 2N contribution to the $K^-$ potential is visible as the difference between the solid and dotted lines. The depth of the absorptive $K^-NN$ potential at $\rho_0$ is $\approx 100$~MeV. The middle panels show the $K^-N$ absorptive potential (dotted), the $t\rho$ potential (dashed), and the total $K^-$ potential ($t\rho+V_{K^-NN}^{\rm corr}$, solid) according to Eq.~\eqref{eq:V_K_YNKpi} calculated with the Pauli + YNK amplitudes. The $t\rho$ potential now includes 1N processes, 2N processes corresponding to the 2FL diagram (d1) in Fig.~\ref{fig:5}, and to some extent also 3N, 4N, ... processes. The difference between the $t\rho$ and the total potential is due to the 2N correction term $V_{K^-NN}^{\rm corr}$. This term includes contributions from the 2FL diagram (d2) and 1FL diagrams (e1) - (e4) in Fig.~\ref{fig:5}. The dominant contribution from the 2FL diagram (mainly the diagram with $\pi\pi$ exchange) causes the imaginary part of the correction term to be negative. The multinucleon absorptive potential (the difference between the solid and dotted lines) is now about $\approx 60$~MeV deep at $\rho_0$. The bottom panels show the same contributions to the total $K^-$ potential as the middle panels but for the Pauli + YNK$\pi$ amplitude model. 
In this case, the correction term $V_{K^-NN}^{\rm{corr}}$ modifies the real part of the potential only slightly but has a positive imaginary part, reducing the absorptive part of the $t\rho$ potential by about $\approx 7$~MeV at $\rho_0$.
This is because, besides the contribution from 1FL diagrams, the term  $V_{K^-NN}^{\rm corr}$ includes the contribution from 2FL diagram (d2) covering only $\eta\eta$ and $\eta\pi$ exchange in this case. The coupling of the $\eta N N$ vertex is about 4 times weaker than the $\pi N N$ one. Therefore, the contribution from the 2FL diagram is very small  and the overall correction term is  positive. The depth of the absorptive multinucleon potential (the difference between the dotted and solid lines) is $\approx 45$~MeV at $\rho_0$.

A chiral model with hadron self-energies was originally developed in the year 2000~\cite{ramos:2000npa}. In the bottom panel of Fig. \ref{fig:7}, we present a comparison of the $K^-$ potential obtained in that work (denoted as '$t\rho$ 2000', dashed black) with the current $K^-$ potential (denoted as '$t\rho$ 2025', dashed magenta) as functions of the relative density $\rho/\rho_0$. It is to be noted that the $t\rho$ potential from 2000 also contains, by the same construction, the contribution from 2N processes corresponding to diagrams (d1) and (d2) depicted in Fig.~\ref{fig:5} with $\bar{K}$ and $\pi$ exchange, respectively, and also contributions from 3N, 4N, ... processes. The difference between the $t\rho$ potentials is due to the fact that, in the 2000 version, only the leading-order (Weinberg-Tomozawa) amplitudes were used, while, in the present work, we used next-to-leading order amplitudes of the BCN model~\cite{feijoo:2019prc}. The BCN amplitudes yield a more attractive real part of the $K^-$ potential at low densities, $\rho < 0.4 \rho_0$, and smaller depth of Re$V_{K^-}\approx-30$~MeV at $\rho_0$ than the model from Ref.~\cite{ramos:2000npa}. The imaginary parts of $V_{K^-}$ are identical at very low densities, $\rho < 0.1\rho_0$, while the BCN model is more absorptive at saturation density than the old potential, reaching a depth of Im$V_{K^-}(\rho_0)\approx-60$~MeV. 

We next applied the model for the $K^-$ potential in calculations of energy shifts and widths in kaonic atoms. We performed calculations for 24 nuclear species and 64 data points from bubble chamber experiments. In Table~\ref{tab:chi2_atoms}, we present the values of $\chi^2$ resulting from comparison with the experimental data \cite{friedman:1994npa} for calculations using the Pauli, Pauli+YN, Pauli+YNK, Pauli+YNK$\pi$ chiral amplitudes and $K^-N$+phen. multiN potential. We should stress that our calculations do not contain any free parameter fitted to kaonic atom data. The values of $\chi^2$ show remarkable improvement when the hadron self-energies are included into the model. The full model, Pauli+YNK$\pi$, yields value of $\chi^2/d.p.$ of 1.5, which is even better than the result for the best fit $K^-N$+phen. multiN potential based on the BCN amplitudes, $\chi^2/d.p.=$1.9. It is worth mentioning that the calculations were performed using the RMF density distributions while the parameters of the best fit $K^-N$+phen. multiN potential were fitted to the data using 2-parameter Fermi distributions for densities \cite{friedman:1994npa}.
\begin{table}[t]
 \caption{Values of $\chi^2(64)$ obtained in calculations of kaonic atoms (24 nuclear species, 64 data points) using the $K^-N + K^-NN$ potentials based on the Pauli, Pauli+YN, Pauli+YNK, and Pauli+YNK$\pi$ BCN amplitudes. For comparison, there are also results of calculations with the $K^-N$+phen. multiN potential based on WRW modified BCN amplitudes.} \vspace{10pt}
 \begin{tabular}{lccccc} \label{tab:chi2_atoms}
BCN & Pauli & ~Pauli+YN~ & ~Pauli+YNK  &  ~{\bf Pauli\mbox{\boldmath$+$}\!YNK}\mbox{\boldmath$\pi$}~ & ~WRW~   \\ \hline \hline
    & ~$K^-N\!+\!K^-NN$~ & ~$K^-N\!+\!K^-NN$~ & ~$t\rho \! + \!V_{K^-NN}^{corr}$~ & ~ ${\bf t \mbox{\boldmath$\rho$}\!\mbox{\boldmath$+$} \!V_{K^-NN}^{corr}}$~ & ~$K^-N$+phen. multiN~ \\ \hline
  $\chi^2 (64)$  &  553.2  &265.5  & 169.2  & {\bf 96.0} & 122.0  \\[1ex]
 $\chi^2$/d.p.  &  8.6 &4.2 &  2.6 & {\bf 1.5} & 1.9 \\ \hline
\end{tabular}
\end{table} 
It is to be noted that A. Baca et al.~\cite{bacaNPA2000} used the microscopic optical potential from Ref.~\cite{ramos:2000npa} that already included the effect of hadron self-energies in calculations of energy shifts and widths in kaonic atoms. They obtained a fairly good value of $\chi^2/d.p.=3.8$. The method of incorporating the hadron self-energies into the chiral amplitudes in our current work is the same as in Ref.~\cite{ramos:2000npa}. The main difference is that we use the next-to-leading order version of the Barcelona model \cite{feijoo:2019prc} and we include the correction term $V_{K^-NN}^{\rm corr}$ for the two-nucleon absorption process on top of the $t\rho$ potential (see Fig.~\ref{fig:7}). 

In Table~\ref{tab:atoms}, we present the values of energy shifts $\Delta\epsilon$ and widths $\Gamma$ for lower and $\Gamma^*$ for upper levels in 24 nuclear species, calculated with the full model for the $K^-$ potential based on the Pauli+YNK$\pi$ BCN amplitudes. Our calculations are compared with the corresponding experimental values from Ref.~\cite{friedman:1994npa} with their uncertainties listed in parentheses. In the case of multiple measurements for a given transition we present the weighted average value determined by C. Batty and E. Friedman \cite{friedman:private}. Most of the calculated values lie within the errors of experimental data and give rise to $\chi^2/d.p.=1.5$. 
\begin{table}[htp]
 \caption{Energy shifts $\Delta\epsilon$ and lower and upper level widths $\Gamma$ and $\Gamma^*$, respectively, calculated using the $K^-$ potential based on the Pauli+YNK$\pi$ BCN amplitudes in nuclei across the periodic table. Experimental values of the corresponding energy shifts and widths with their uncertainties (in parentheses) are presented for comparison.}
    \begin{tabular}{c|c|ccc|m{2.6cm}m{2.6cm}m{2.6cm}} 
        Nucleus & Level & \multicolumn{3}{c|}{Pauli+YNK$\pi$} & \multicolumn{3}{c}{Exp.~\cite{friedman:1994npa}}  \\ \hline \hline
       &  & $\Delta\epsilon$ (keV) &  $\Gamma$ (keV) & $\Gamma^*$ (eV)  & $\Delta\epsilon$ (keV) & $\Gamma$ (keV) & $\Gamma^*$ (eV) \\ \hline
$^7$Li & $2p$ &  -0.003     &  0.054  & -  & 0.002 (0.026)   & 0.055 (0.029)  & - \\ 
$^9$Be & $2p$ &   -0.037  &    0.247 &   0.03  &  -0.079 (0.021) & 0.172 (0.058) &  0.04 (0.02) \\ 
$^{10}$B & $2p$&   -0.170  &  0.72 &  - &   -0.208 (0.035)&   0.81 (0.10) &    -  \\   
$^{11}$B & $2p$ &  -0.179  &   0.76 &    -  &  -0.167 (0.035)&   0.70 (0.08)&    - 
 \\  
 $^{12}$C & $2p$ &  -0.54 &    1.72 &    0.62 &  -0.59 (0.08)&  1.73 (0.15) &   0.99 (0.20)  \\ 
$^{16}$O & $3d$ &  -0.0004&  0.008   & - &  -0.025 (0.018)&   0.017 (0.014)  &    -  \\  
$^{24}$Mg & $3d$ &  -0.036 &   0.220 &  0.13 &  -0.027 (0.015) &   0.214 (0.015)  & 0.08 (0.03)  \\ 
$^{27}$Al & $3d$ &  -0.077 &   0.418 &  0.33  &  -0.080 (0.013)  &   0.443 (0.022) & 0.30 (0.04)  \\ 
$^{28}$Si & $3d$ &-0.147   &   0.677 &  0.65 &  -0.139 (0.014)&   0.801 (0.032) & 0.53 (0.06)  \\ 
$^{31}$P & $3d$  & -0.32&   1.39 &   1.93 &  -0.33 (0.08)  &   1.44 (0.12) & 1.89 (0.30)  \\ 
 $^{32}$S & $3d$ & -0.55 &   1.92  &  3.15 &  -0.494 (0.038)&   2.19 (0.10)  & 3.03 (0.44) \\ 
$^{35}$Cl & $3d$ & -0.97 &   3.00  &  6.3 &  -0.99 (0.17)  &   2.91 (0.24) & 5.8 (1.7)  \\ 
$^{59}$Co & $4f$ & -0.115 &   0.64  &   - &  -0.099 (0.106) &   0.64 (0.25)  & -  \\  
$^{59}$Ni & $4f$ &  -0.165  &   0.82 &  2.2 &  -0.223 (0.042) &   1.03 (0.12)& 5.9 (2.3)  \\  
 $^{63}$Cu & $4f$ &  -0.26 &   1.20 &  3.7 &  -0.370 (0.047)&   1.37 (0.17) & 5.2 (1.1)  \\ 
$^{108}$Ag & $4f$ &  -0.24 &   1.37 & 8.3 &  -0.18 (0.12) &   1.54 (0.58) & 7.3 (4.7) \\ 
$^{112}$Cd & $5g$ &   -0.33 &   1.82 &  12.2  &  -0.40 (0.10) &   2.01 (0.44) & 6.2 (2.8) \\
$^{115}$In & $5g$ & -0.43&   2.34   &  17.2 &  -0.53 (0.15)  &   2.38 (0.57)  & 11.4 (3.7)  \\ 
$^{118}$Sn & $5g$ & -0.56  &   2.96 &  23.7 &  -0.41 (0.18)  &   3.18 (0.64) & 15.1 (4.4)  \\ 
 $^{165}$Ho & $6h$ &  -0.17 &   1.24&   - &  -0.30 (0.13) &   2.14 (0.31)   & -  \\
$^{173}$Yb & $6h$ &   -0.34 &   2.18 &   -  &  -0.12 (0.10)&   2.39 (0.30) & -  \\  
 $^{181}$Ta & $6h$ & -0.64 &   3.55 &   - &  -0.27 (0.50)  &   3.76 (1.15) & -  \\ 
$^{208}$Pb & $7i$ &  -0.01 &   0.28 &  2.60 &  -0.020 (0.012) &   0.37 (0.15) &  4.1 (2.0)  \\ 
 $^{238}$U & $7i$ & -0.15  &   1.58  &  22 &  -0.26 (0.4) &   1.50 (0.75) &  45 (24)  \\ \hline
 \end{tabular}
    \label{tab:atoms}
\end{table}

\begin{figure}[t]
    \centering
    \includegraphics[width=0.43\textwidth]{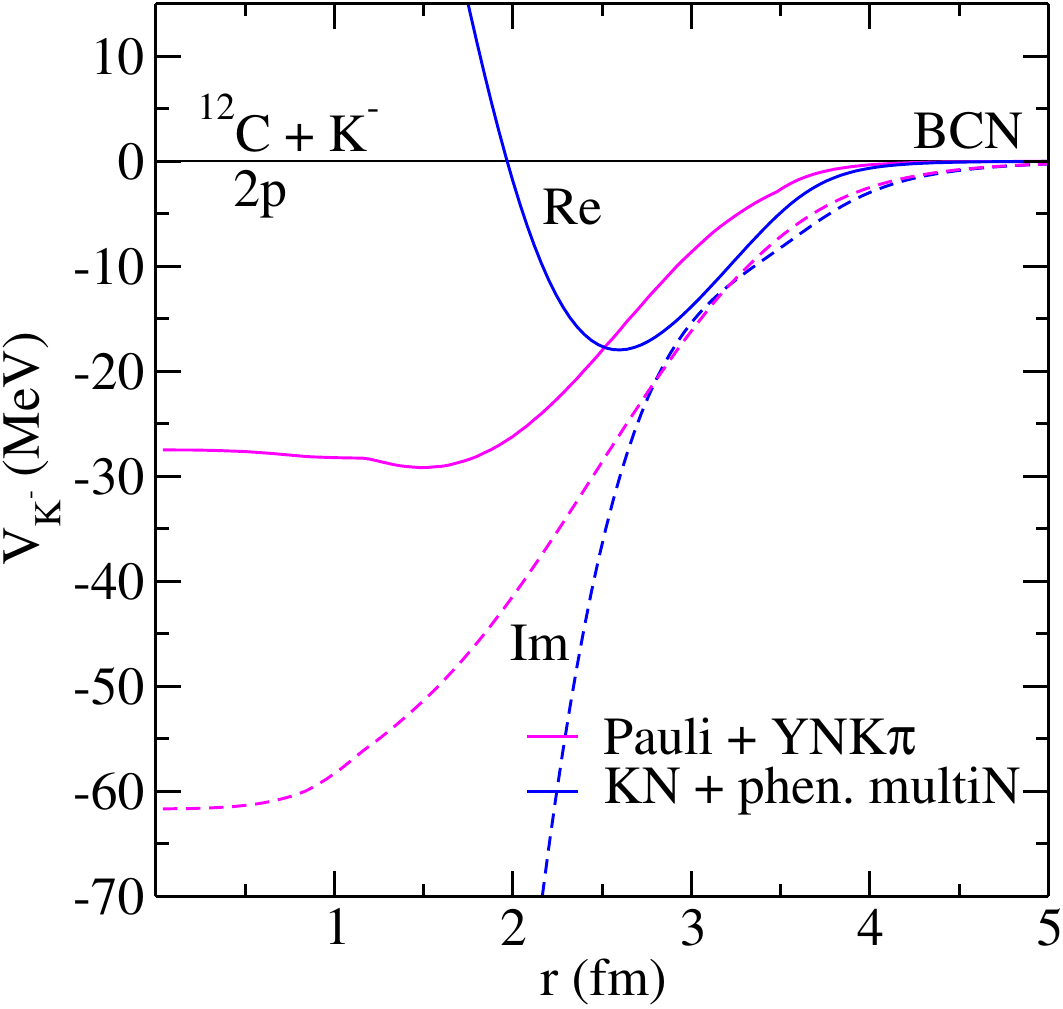} \hspace{5pt}
    \includegraphics[width=0.44\textwidth]{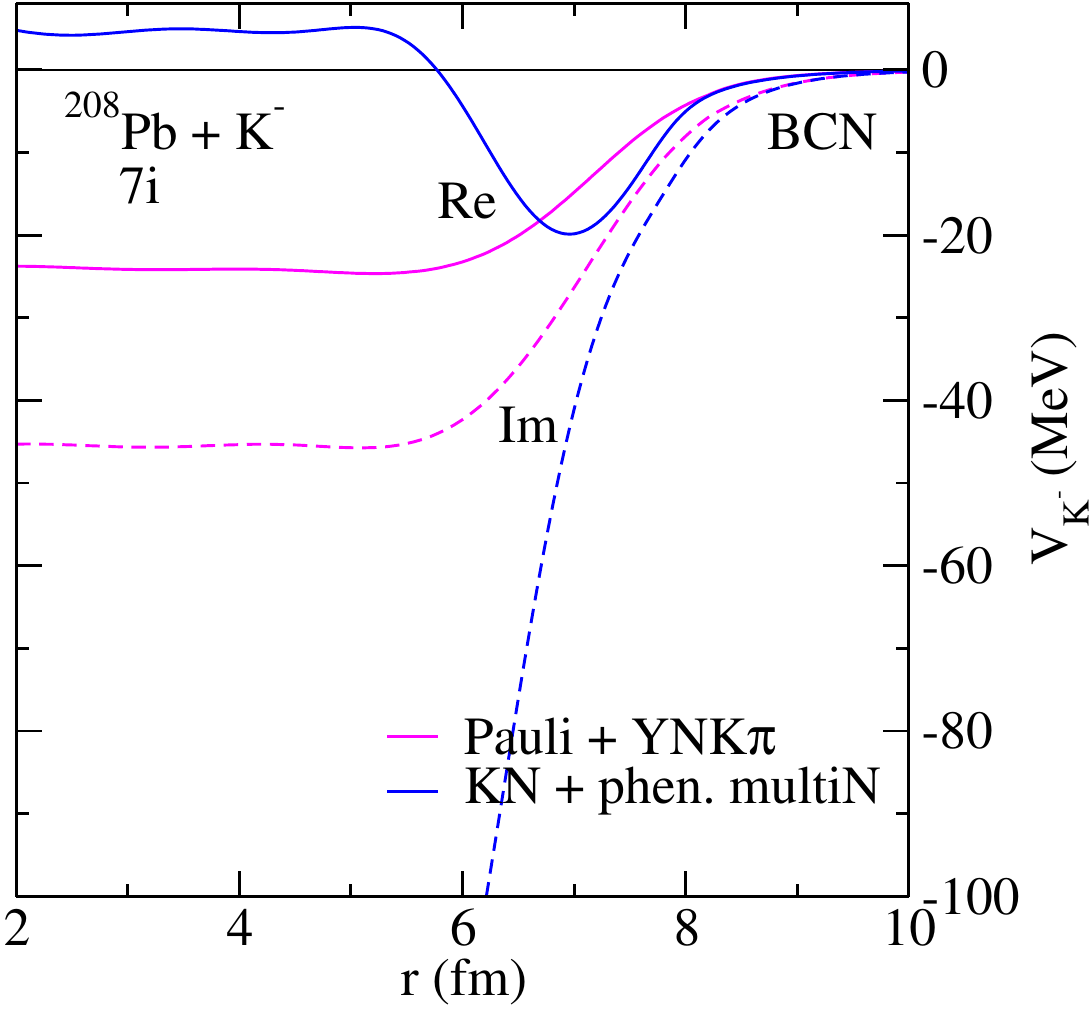} 
    
    \vspace{10pt}
    
    \includegraphics[width=0.91\textwidth]{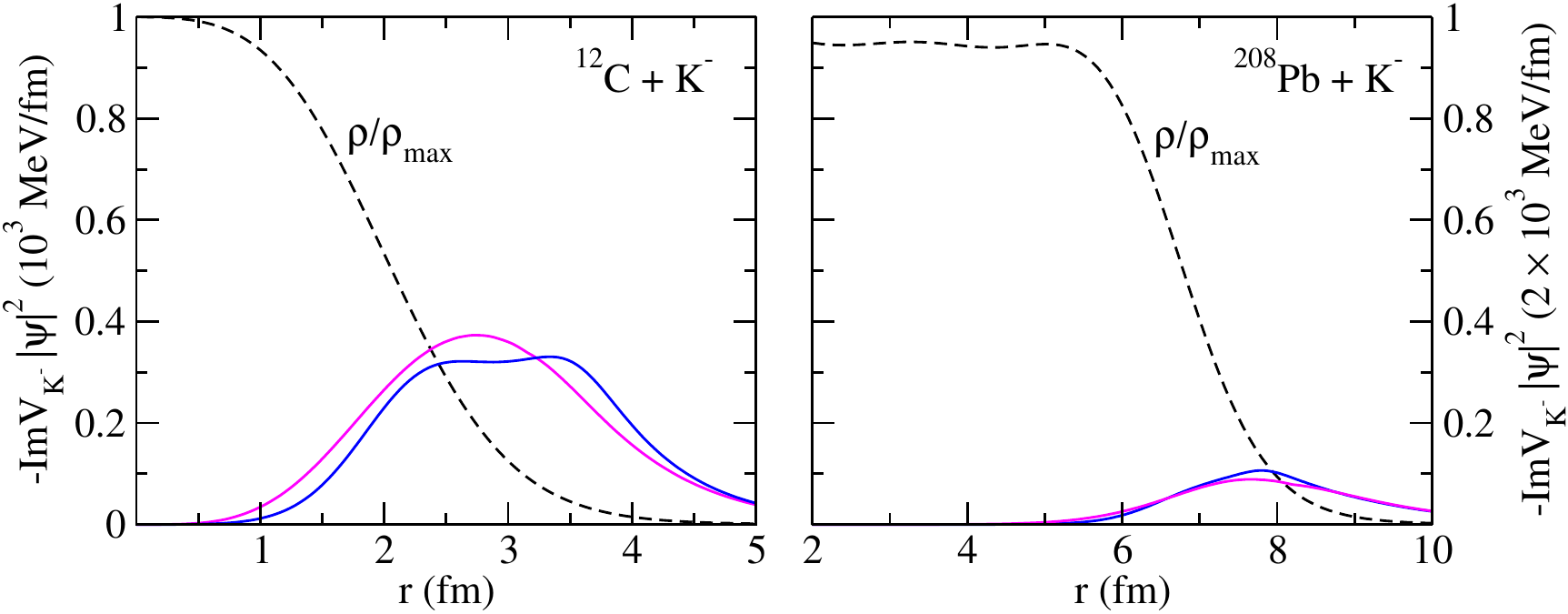}
    \caption{Total $K^-$ potential in the $^{12}$C+$K^-$ (left) and $^{208}$Pb+$K^-$ (right) atoms calculated with the Pauli+YNK$\pi$ BCN amplitudes (magenta) compared with the $K^-N$+phen. multiN potential (blue) based on the BCN amplitudes. Lower panel shows the corresponding overlap of Im$V_{K^-}$ with the $K^-$ wave function squared $|\psi|^2$. The relative density $\rho/\rho_{\rm max}$ is shown for illustration.}
    \label{fig:8}
\end{figure}

Figure~\ref{fig:8} shows the real (solid) and imaginary (dashed) parts of the total $K^-$ optical potential as a function of the nuclear radius $r$ in $^{12}$C$+K^-$ (top left) and $^{208}$Pb$+K^-$ (top right). The potentials were calculated using the Pauli+YNK$\pi$ BCN amplitudes (magenta). The best fit $K^-N$+phen. multiN potential (blue) is presented for comparison. Lower panels show the overlap of Im$V_{K^-}$ with the $K^-$ wave function squared $|\psi|^2$, i.e., they indicate the region of nuclear density where the absorption of the $K^-$ occurs. 
We can see that the depths of both potentials, microscopic and phenomenological, are very close to each other in the region of maximal overlap of $|\psi|^2$ with Im$V_{K^-}$, i.e., at the nuclear surface. Therefore, they both yield very similar values of energy shifts and widths in agreement with experimental data. On the other hand, the potentials are very different in the nuclear interior. The $K^-N$+phen. multiN potential, multinucleon part of which was fitted to kaonic atom data, predicts repulsive real part and very deep imaginary part reaching $\approx -300$~MeV for $^{12}$C and $\approx -160$~MeV for $^{208}$Pb in the nuclear interior. However, one should keep in mind that kaonic atom data can reliably determine the mutlinucleon potential only up to $\approx 0.5\rho_0$. The full potential model based on Pauli+YNK$\pi$ BCN amplitudes predicts Re$V_{K^-}(r=0)\approx -27$~MeV and Im$V_{K^-}(r=0)\approx -61$~MeV for $^{12}$C and Re$V_{K^-}(r=0)\approx -25$~MeV and Im$V_{K^-}(r=0)\approx -47$~MeV for $^{208}$Pb. Such big differences between the microscopic and phenomenological potentials in the nuclear interior do not affect results for kaonic atoms but would have a dramatic impact for predictions of $K^-$-nuclear bound states \cite{hmPRC2017}.

Next, we calculated branching ratios for mesonic and non-mesonic absorption channels in kaonic carbon ($^{12}$C+$K^-$) and neon ($^{20}$Ne+$K^-$) and compared them with available experimental data on primary-interaction branching ratios, i.e., corrected for secondary interactions of particles created in the absorption process. The branching ratios were evaluated as fractions of the partial width in the respective channel (see Table~\ref{tab:channels}) over the total width
\begin{equation} \label{eq:br_ratio}
\text{BR} = \frac{\Gamma_{\rm channel}}{\Gamma_{\rm total}} = \frac{\int \text{Im}V_{\rm channel}(r) |\psi(r)|^2 dr }{\int \text{Im}V_{K^-}(r)|\psi(r)|^2 dr }~,    
\end{equation}
where $\psi(r)$ is the $K^-$ radial wave function. 

In Table~\ref{tab:ratios_C}, we show mesonic and non-mesonic branching ratios in the $^{12}$C$+K^-$ atom ($3d$ level), calculated with the total $K^-$ potential based on the Pauli + YN, Pauli+YNK, and Pauli+YNK$\pi$ BCN amplitudes. The $K^-$ potential for single-nucleon channels was calculated from 1N diagrams of the type shown in Fig.~\ref{fig:4}. In the case of the Pauli + YN model, we present the total 2N absorption ratio originating from all diagrams in Fig.~\ref{fig:5}. For the rest of the models we show the total multinucleon absorption ratio.
\begin{table}[t]
 \caption{Primary-interaction branching ratios (in $\%$) for mesonic ($K^-N\rightarrow Y \pi$, $Y=\Lambda, \Sigma$) and non-mesonic absorption of $K^-$ in the $^{12}$C$+K^-$ atom (3d level), calculated with the total $K^-$ potentials based on the Pauli + YN, Pauli + YN$K^-$, and Pauli + YN$K^-\pi$ BCN amplitudes. The experimental data for primary-interaction branching ratios are shown for comparison.}
 \begin{tabular}{l|ccc||c|c} \label{tab:ratios_C}
$^{12}$C + $K^-$ ($3d$)  & \multicolumn{3}{c||}{BCN}   & \multicolumn{2}{c}{Exp.}   \\ \hline
mesonic ratio  & ~Pauli + YN~ & ~Pauli+YNK~ & ~Pauli+YNK$\pi$~  &~\cite{vander:1977nc}~ & ~\cite{davis:1968nc}~  \\ \hline \hline
$\Sigma^+ \pi^- $ & 17.0  &15.8  & 14.4  & 29.4 $\pm$ 1.0 & 14.4 $\pm$ 2.3\\
$\Sigma^- \pi^0 $ &  8.7 &8.3  & 8.1  &2.6 $\pm$ 0.6 &  1.2 $\pm$ 0.4 \\
$\Sigma^- \pi^+ $ & 22.3  &14.1 & 12.8  & 13.1 $\pm$ 0.4 &  10.3 $\pm$ 1.7\\
$\Sigma^0 \pi^-$ &  8.8  &8.4 & 8.2  & 2.6 $\pm$ 0.6 &  1.2 $\pm$ 0.4 \\
$\Sigma^0 \pi^0 $  & 16.4 &11.4  & 10.0  & 20.0 $\pm$ 0.7 &  11.8 $\pm$ 1.4 \\
$\Lambda \pi^0 $  &  5.3  &5.3  & 5.3  & 3.4 $\pm$ 0.2 & 11.8 $\pm$ 1.0 \\ 
$\Lambda \pi^-$ &  11.5 &11.2  & 11.1  & 6.8 $\pm$ 0.3  & 23.6 $\pm$ 1.9  \\ \hline
total 1N ratio& 90.0   &74.3  & 70.0  &77.9 $\pm$ 1.6 & 74.3 $\pm$ 3.9 \\ \midrule 
\multirow{2}*{total multiN ratio} & \multirow{2}*{10.0\footnote{2N ratio} } & \multirow{2}*{25.7}  & \multirow{2}*{30.0} & 19.0 $\pm$ 2.0 & 25.7 $\pm$ 3.1   \\
 & & & & \multicolumn{2}{c}{27 $\pm$ 3(stat.)$^{+5}_{-6}$(syst.) \cite{amadeus2020} }\\
\end{tabular}
\end{table} 
The experimental data on primary-interaction branching ratios are taken from bubble chamber experiments: $K^-$ absorption in $^{12}$C \cite{vander:1977nc} and $K^-$ absorptions in a mixture of 76\% CF$_3$Br + 24\% C$_3$H$_8$ \cite{davis:1968nc}.  
According to cascade model calculations \cite{manti:private} about 70\% of all absorptions in kaonic carbon occur in the $3d$ level and 10\% in the $4f$ one. Therefore, we compare experimental data with branching ratios calculated for the $3d$ level. All three amplitude models yield very similar branching ratios for $\Sigma^- 
\pi^0, \Sigma^0 \pi^-, \Lambda \pi^0$, and $\Lambda \pi^-$ channels. On the other hand, the experimental values for these channels are much lower, except the value for the $\Lambda \pi^0$ ratio from Ref.~\cite{davis:1968nc}. The most striking difference is between the $\Sigma^+ \pi^-, \Sigma^- \pi^+$, and $\Sigma^0 \pi^0$ absorption channels. The Pauli + YN model yields the largest values of branching ratios for these channels. The ratios gradually decrease when the $K^-$ and $\pi$ self-energies are considered and in the final Pauli + YNK$\pi$ model they become consistent with the data from Ref.~\cite{davis:1968nc}. It is to be noted that the experimental values for these channels differ considerably between each other. For instance, the values of fractions for the $\Sigma^+ \pi^-$ and $\Sigma^0 \pi^0$ channels in Ref.~\cite{vander:1977nc} are about twice larger than the corresponding values in Ref.~\cite{davis:1968nc}. The experimental data are very sensitive to corrections due to the final state effects such as $\Sigma N - \Lambda N$ conversion probability. It seems that the experimental works implemented the $\Sigma N - \Lambda N$ conversion differently.  Indeed, in Ref.~\cite{vander:1977nc},a conversion probability of $64\%\pm3\%$ for $\Sigma^+$ and $62\%\pm2\%$ for $\Sigma^-$ was used to obtain the primary-interaction ratios. In Ref.~\cite{davis:1968nc}, the primary-interaction ratios were evaluated by assuming a 45\% conversion probability for $\Sigma^+$ and $\Sigma^0$ and a $50\%$ conversion probability for $\Sigma^-$.  In the earlier work of Vander Velde-Wilquet et al.~\cite{vander:1975npa}, the conversion probability for $\Sigma^-$ on proton was found to be $38\%\pm3\%$ for $^{12}$C and the corresponding primary-interaction branching ratio for $\Sigma^- \pi^+$ was established to 12.7\% and for $\Sigma^+ \pi^-$ it was $15.9\%$. The same data were reanalyzed in Ref. \cite{vander:1977nc} and about twice larger values for $\Sigma^+ \pi^-$ primary branching ratio was established (see Table~\ref{tab:ratios_C}). 

If we compare the sum of experimental mesonic branching ratios having a $\pi^-$ in the final state we get $38.8\% \pm 1.2\%$ and $39.2\% \pm 3.0\%$, employing the data of \cite{vander:1977nc} and \cite{davis:1968nc}, respectively. Likewise, for $\pi^0$ in the final state we have $26.0\% \pm 0.9\%$ and $24.8\% \pm 1.8\%$, respectively.  Therefore, both experiments are compatible with each other. The sum of branching ratios with $\pi^-$ in  the final state for the Pauli + YNK model is 35.4\% and for the Pauli + YNK$\pi$ model it is 33.7\%, which is somewhat lower than experimental value. For channels with $\pi^0$ in the final state the sum is 25.0\% for the Pauli + YNK model and 23.4\% for the Pauli + YNK$\pi$ one, in fair agreement with the experiment. In this later case, the resulting single-nucleon absorption ratio is 70\% while the corresponding multinucleon ratio is 30\%. If we calculate weighted average for the $3d+4f$ levels, the 1N ratio becomes 71.8\% and the mutlinucleon one 28.2\%, which lies within the uncertainties of the experimental values from Ref.~\cite{davis:1968nc} and from the AMADEUS collaboration \cite{amadeus2020}. 

\begin{table}[t]
 \caption{Primary-interaction branching ratios (in $\%$) for mesonic ($K^-N\rightarrow Y \pi$, $Y=\Lambda, \Sigma$) and non-mesonic absorption of $K^-$ in the $^{20}$Ne$+K^-$ atom ($3d$ level), calculated with the $K^-$ potentials based on the Pauli + YN, Pauli + YNK, and Pauli + YNK$\pi$ BCN amplitudes. The experimental data for primary-interaction branching ratios are shown for comparison.}
 \begin{tabular}{l|ccc||c} \label{tab:ratios_Ne}
$^{20}$Ne + $K^-$ ($3d$)  & \multicolumn{3}{c||}{BCN}   & Exp.  \\ \hline
mesonic ratio  & ~Pauli+YN~ &  ~Pauli+YNK~  & ~Pauli+YNK$\pi$~  &~\cite{moulder:1971npb}~ \\ \hline \hline
$\Sigma^+ \pi^- $ & 17.6  & 15.8 & 14.0  & 29.2 $\pm$ 2.9 \\
$\Sigma^- \pi^0 $ & 8.0  &  7.3 & 7.0   &1.6 $\pm$ 0.7  \\
$\Sigma^- \pi^+ $ & 22.4  & 12.7 & 10.8  & 15.8 $\pm$ 2.4\\
$\Sigma^0 \pi^-$ & 8.1 & 7.4 & 7.1 & 1.6 $\pm$ 0.7 \\
$\Sigma^0 \pi^0 $ & 16.9 & 10.9  & 9.2 & 20.7 $\pm$ 1.8 \\
$\Lambda \pi^0 $ & 5.2  & 5.1  & 5.0 & 2.6 $\pm$ 0.9 \\ 
$\Lambda \pi^-$ & 10.7 & 10.2 & 9.9  & 5.2 $\pm$ 1.3  \\ \hline
total 1N ratio & 88.9& 69.3 & 63.0 &76.7 $\pm$ 4.6  \\ \midrule 
\multirow{2}*{total multiN ratio} & \multirow{2}*{11.1\footnote{2N ratio}} & \multirow{2}*{30.7} & \multirow{2}*{37.0}  & 22.6 $\pm$ 3.0  \\
  &  &  &  & ~27.5 $\pm$ 3.0\footnote{Assuming no pion reabsorption}
\end{tabular}
\end{table} 
In Table~\ref{tab:ratios_Ne}, we show a comparison of the absorption fractions from the $3d$ level in $^{20}$Ne+$K^-$, obtained within the Pauli + YN, Pauli + YNK and Pauli + YNK$\pi$ models, with experimental data of primary-interaction branching ratios \cite{moulder:1971npb}. Cascade calculations show that about 73\% of all absorptions occur at the $3d$ level and about 18\% at the $4f$ level in kaonic neon. The calculated fractions are very similar to those in kaonic carbon. The values of the $\Sigma^+ \pi^-, \Sigma^- \pi^+$, and $\Sigma^0 \pi^0$ ratios become again smaller when the kaon and pion self-energies are considered in comparison with the Pauli + YN model. The experimental values for the $\Sigma^+ \pi^-$ and $\Sigma^0 \pi^0$ ratios are twice as large as in the Pauli + YNK and Pauli + YNK$\pi$ models, similarly to what has been observed for $^{12}$C in Table~\ref{tab:ratios_C}. Note that in Ref.~\cite{moulder:1971npb} the $\Sigma^-$ conversion probability was established to be $51\% \pm 8$\%, that of the $\Sigma^+$ was $70\% \pm2$\%, and for $\Sigma^0$ a value of $60\% \pm 16$\% was found. These values were used to obtain the primary-interaction ratios presented in the last column of Table~\ref{tab:ratios_Ne}.
As before, when we sum up the channels with $\pi^-$ in the final state, we obtain the experimental value of $36.0\% \pm 3.3\%$, which is comparable with the value of 33.4\% for Pauli + YNK. On the other hand, the value of 31.0\% for the Pauli + YNK$\pi$ model is again too low. For channels with $\pi^0$ in the final state, the Pauli + YNK model yields value of 23.3\% which is again in agreement with the sum of experimental ratios, $24.90\% \pm 2.13\%$. The Pauli + YNK$\pi$ model gives a value of 21.2\% which is again a bit lower than the experiment. The total 1N ratio in the Pauli + YNK$\pi$ model is 63\% and the corresponding multinucleon ratio is 37.0\%. With a weighted average for $3d+4f$ levels, the single-nucleon absorption ratio increases slightly to 65\% while the corresponding multinucleon ratio decreases to 35\%, these values are still a bit off the experimental data shown in Table~\ref{tab:ratios_Ne}.  

The FINUDA collaboration measured the emission rates in $K^- A \rightarrow \pi^{\pm} \Sigma^{\mp} A^{\prime}$ reactions in p-shell nuclei \cite{finuda2011}. Their data are, however, not corrected for pion attenutation and $\Sigma N - \Lambda N$ conversion. In order to compare their results with our calculation we have to correct the data for the two effects mentioned above. We did it according to the following expression
\begin{equation}
    BR_{\rm primary} = \frac{BR_{\rm exp}}{(1-P_{\pi})(1-P_{\Sigma-\Lambda})},
\end{equation}
where $P_{\pi}$ stands for the pion absorption probability and $P_{\Sigma-\Lambda}$ denotes the $\Sigma N- \Lambda N$ conversion probability. For the probability of pion absorption in carbon we take the value of 8\% \cite{vander:1977nc}. Since the pion absorption depends on the number of nucleons we estimate the absorption probability per nucleon from the value for carbon and then multiply it by the number of nucleons for each considered nucleus. For the $\Sigma N - \Lambda N$ conversion probability we consider two scenarios: a) 60 \% conversion probability for $\Sigma^+$ and 22.5 \% for $\Sigma^-$ \cite{katz:1970prd} b) 50 \% conversion probability for both $\Sigma$'s \cite{moulder:1971npb}. The corrected experimental absorption ratios compared with the corresponding theoretical values, calculated using the Pauli + YN, Pauli + YNK, and Pauli + YNK$\pi$ model, are presented in Table~\ref{tab:br_finuda_pn}.
\begin{table}[t]
\caption{Primary-interaction ratios (in \%) for the $\Sigma^- \pi^+$ and $\Sigma^+ \pi^-$ channels in light kaonic atoms calculated using the Pauli + YN, Pauli + YNK, and Pauli+YNK$\pi$ BCN amplitudes. Experimental data from the FINUDA collaboration \cite{finuda2011} are corrected for pion attenuation and for $\Sigma N- \Lambda N$ conversion with probabilities (a) 60 \% for $\Sigma^+$ and 22.5 \% for $\Sigma^-$ \cite{katz:1970prd} (b) 50 \% for both $\Sigma$'s \cite{moulder:1971npb}. In this calculation, realistic proton and neutron densities were used, instead of $\rho_p=\rho_n=\rho/2$ in the $K^-N$ potential in Eq.~\eqref{eq:imVknpiY+YN_se}.}
\begin{tabular}{c|cc|cc|cc||c|c|c|c} \label{tab:br_finuda_pn}
      & \multicolumn{6}{c||}{BCN} & \multicolumn{4}{c}{Exp.~\cite{finuda2011}} \\ \hline
      & \multicolumn{2}{c|}{Pauli+YN} & \multicolumn{2}{c|}{Pauli+YNK} & \multicolumn{2}{c||}{Pauli+YNK$\pi$}  & \multicolumn{2}{c|}{(a)}  &  \multicolumn{2}{c}{(b)} \\ \hline
    nucleus ($nl$) & $\Sigma^+ \pi^-$ & $\Sigma^- \pi^+$& $\Sigma^+ \pi^-$ & $\Sigma^- \pi^+$&  $\Sigma^+ \pi^-$ & $\Sigma^- \pi^+$ & $\Sigma^+ \pi^-$ & $\Sigma^- \pi^+$ & $\Sigma^+ \pi^-$  & $\Sigma^- \pi^+$ \\ \hline

    $^{6}$Li ($2p+3d$) & 18.8 & 22.5 &17.3 & 18.5& 16.1  & 17.5   &  45.8 $\pm$ 4.3  & 19.4 $\pm$ 1.5 &  36.7 $\pm$ 3.4   & 30.0 $\pm$ 2.3 \\
    $^{7}$Li ($2p+3d$) & 16.3 & 16.8 &14.1 & 12.1&12.6  &  11.0   &  20.5 $\pm$ 1.4  &  8.7 $\pm$ 1.0 &  16.4 $\pm$ 1.1    &  13.4 $\pm$ 1.5\\
    $^{9}$Be ($3d$)& 15.7&18.9 &14.3 &13.6 & 13.1 &   12.6  & 14.6 $\pm$ 1.1  & 5.9 $\pm$ 0.7 & 11.7 $\pm$ 0.9    &  9.2 $\pm$ 1.2 \\
    $^{12}$C\footnote{Branching ratios for $^{12}$C were measured in bubble chamber experiment \cite{vander:1977nc, vander:1975npa} but they served as reference values in Table 1 in Ref.~\cite{finuda2011}.} ($3d$)&17.2 & 22.7 &16.1 & 14.5& 14.7 &   13.2  &  25.0 $\pm$ 0.8  &  10.2 $\pm$ 0.3 & 20.0 $\pm$ 0.7    &  15.9 $\pm$ 0.4\\
    $^{13}$C ($3d$) & 15.8 & 19.0 & 14.2 & 11.6& 12.7  & 10.3  & 18.6 $\pm$ 2.7  &  6.5 $\pm$ 0.8 &  14.9 $\pm$ 2.2   &  10.1 $\pm$ 1.2 \\
    $^{16}$O ($3d$)&18.0 &22.4 &16.1 &13.2 & 14.5 &  11.6  &  19.3 $\pm$ 1.9  & 8.4 $\pm$ 0.8 & 15.5 $\pm$ 1.5    &  13.0 $\pm$ 1.2 \\
    
\end{tabular}

\end{table}
The ratios for Li were calculated as a weighted average of the $2p+3d$ levels. According to cascade model calculations, these levels have the largest probability of kaon absorption, namely 20\%(25\%) for $2p$($3d$) level in $^{6}$Li and 33\%(26\%) for $2p$($3d$) level in $^{7}$Li. In the rest of the nuclei, the largest absorption  probability is for the $3d$ level, namely 50\% for $^9$Be, 70\% for $^{12}$C, and 80\% for $^{16}$O. We assume that the corresponding value is valid also for $^{13}$C. The corrected experimental primary-interaction branching ratios differ quite significantly for the two scenarios of $\Sigma N - \Lambda N$ conversion. This demonstrates great sensitivity of the ratios to the particular choice of this probability. The theoretical calculations with the full model Pauli+YNK$\pi$ are closer to the experimental ratios corrected  for $\Sigma N - \Lambda N$ conversion with scenario (b), except Li\footnote{For $\Sigma N - \Lambda N$ conversion probabilities in light nuclei, values much lower than $50\%$ were estimated in Ref.~\cite{vander:1975npa}}. It is to be noted that the ratios for Li were calculated with densities derived within the RMF model, which may not describe the structure of Li in sufficient detail. A large difference between the experimental ratios in $^{6}$Li and $^{7}$Li is due to the twice larger emission rates measured in $^{6}$Li with respect to those in  $^{7}$Li (see Table 1 in Ref. \cite{finuda2011}).  For nuclei beyond Li, Table~\ref{tab:br_finuda_pn} shows that the higher is the probability of $K^-$ absorption at the $3d$ level, the better is the agreement between the theoretical and experimental primary-interaction branching ratios [scenario (b)].

\section{Conclusions}
\label{conclusions}

In this work, we obtained the $K^-N$ scattering amplitudes in symmetric nuclear matter,  including Pauli blocking effect, nucleon, hyperon, kaon, and pion in-medium self-energies within the NLO chiral meson-baryon coupled-channel interaction model, denoted here as the BCN model \cite{feijoo:2019prc}. The amplitudes were used to derive the $K^-$-nuclear potential, based on a microscopic model originally developed in Ref.~\cite{hrtankova:2020prc}. The microscopic model for the $K^-$ potential has been modified due to the effect of hadron self-energies as well. The $K^-$ potential includes single-, two-, and when the kaons and pions are dressed in the medium also the multinucleon absorption processes.

We performed a systematic study of the influence of the respective hadron self-energies on the energy dependence of the $K^-N$ chiral amplitudes and, consequently, on the depth of the total $K^-$-nuclear potential in symmetric nuclear matter. The full model, including Pauli correlations and all considered hadron self-energies, yielded a relatively shallow $K^-$ potential at saturation density, with Re$V_{K^-}(\rho_0)=-30$~MeV and Im$V_{K^-}(\rho_0)=-50$~MeV. A similar depth for the $K^-$ potential was obtained with the leading-order OR model~\cite{ramos:2000npa}. Such shallow potentials are in contradiction with a purely phenomenological $K^-$ potential fitted to kaonic atom data with depth of Re$V_{K^-} \approx 200$~MeV in the nuclear interior \cite{friedman:1994npa}. However, it should be stressed that kaonic atom data can reliably determine the $K^-$ potential only up to $\approx 50\%$ of $\rho_0$ and they are insensitive to the form of $K^-$ potential at higher densities \cite{fgNPA2017}. Therefore, the large differences between the microscopic and phenomenological potentials in the nuclear interior do not affect the results of kaonic atoms characteristics but would have a dramatic impact for predictions of $K^-$-nuclear bound states. 

We applied the model for the $K^-$ potential in nuclear medium in calculations of strong energy shifts and widths in kaonic atoms. Description of kaonic atom data based solely on a microscopic $K^-$-nuclear potential constructed within the chiral coupled-channel meson-baryon interaction model has been challenging for a long time. In this work, for the first time, we reproduced satisfactorily the 64 data points with a microscopic $K^-$ potential, reaching the value $\chi^2/d.p = 1.5$, which is comparable with $\chi^2/d.p.=1.9$ for the best fit $K^-N$ + phenomenological multinucleon potential based on the BCN amplitudes. 

Finally, we calculated the single-nucleon and multinucleon branching ratios in kaonic carbon and kaonic neon and compared them with available experimental data. The total 1N and multinucleon fractions calculated with the microscopic model are found to be in rather good agreement with experimental data. The respective single-nucleon branching ratios for different $K^-N \rightarrow \pi Y$ channels are distributed in a different way than the experimental ones. However, the experimental data are very sensitive to corrections due to the final state effects, such as $\Sigma N - \Lambda N$ conversion probability which was treated differently in the analyses of relevant experiments. On the other hand, the sums of calculated branching ratios with $\pi^0$ and $\pi^-$ in the final state reproduce nicely the corresponding sums of experimental ratios.

In summary, we have developed a microscopic model for the $K^-$ potential based on the NLO chiral model for the $K^-N$ scattering amplitudes which describes well available $K^-p$ scattering data, threshold branching ratios and energy shift and width of kaonic hydrogen. After inclusion of all relevant in-medium effects the chiral amplitudes give rise to the total $K^-$-nuclear potential, including multinucleon absorption processes, which is capable of describing the 64 data points on strong energy shifts and widths in kaonic atoms with $\chi^2/d.p.=1.5$. Moreover, the model yields the mesonic and non-mesonic branching ratios in fair agreement with available experimental data.

\begin{acknowledgments}
    We thank Simone Manti for providing us with the absorption probabilities in kaonic atoms calculated in the cascade model. We are grateful to Raffaele Del Grande for useful discussions on experimental branching ratios and Eli Friedman for useful comments on kaonic atom data.  A.R. acknowledges financial support from the Maria de Maeztu Center of Excellence grant CEX2024-001451-M and the project PID2023-147112NB-C21, both funded by MICIU/AEI/10.13039/501100011033 (Spain).

\end{acknowledgments}

\end{document}